\documentclass[showpacs,aps,prb,twocolumn,superscriptaddress,psfrag,amsmath,amssymb,longbibliography]{revtex4-2}

\usepackage[pdftex]{graphicx} 
\usepackage{epstopdf}
\usepackage{verbatim}
\usepackage{amsmath}
\usepackage{color}
\usepackage{subfigure}
\usepackage{amsbsy}
\usepackage{wasysym}
\usepackage{textcomp}
\usepackage{times}
\usepackage{float}
\usepackage{latexsym,amsmath,amssymb,bm,euscript}
\usepackage[colorlinks=true,linkcolor=blue,citecolor=blue,urlcolor=blue]{hyperref}
\usepackage{hyperref}
\usepackage{soul}
\usepackage[normalem]{ulem}
\usepackage{mathrsfs}
\usepackage{lettrine}
\usepackage{xspace}
\usepackage{filecontents}

\begin{document}


\title{Spiral spin liquid resilient to quantization in the frustrated honeycomb antiferromagnet GdZnPO}


\author{Xun Chen}
\thanks{These authors contributed equally to this work}
\affiliation{Wuhan National High Magnetic Field Center and School of Physics, Huazhong University of Science and Technology, 430074 Wuhan, China}
\author{Rui Bian}
\thanks{These authors contributed equally to this work}
\affiliation{School of Physics and Beijing Key Laboratory of Opto{-}electronic Functional Materials \& Micro{-}nano Devices, Renmin University of China, 100872 Beijing, China.}
\author{Yuqian Zhao}
\affiliation{Wuhan National High Magnetic Field Center and School of Physics, Huazhong University of Science and Technology, 430074 Wuhan, China}
\author{Haijun Liao}
\affiliation{Institute of Physics, Chinese Academy of Sciences, 100190 Beijing, China.}
\affiliation{Songshan Lake Materials Laboratory, 523808 Dongguan, Guangdong, China.}
\author{Weiqiang Yu}
\email{wqyu\_phy@ruc.edu.cn}
\affiliation{School of Physics and Beijing Key Laboratory of Opto{-}electronic Functional Materials \& Micro{-}nano Devices, Renmin University of China, 100872 Beijing, China.}
\affiliation{Key Laboratory of Quantum State Construction and Manipulation (Ministry of Education), Renmin University of China, 100872 Beijing, China.}
\author{Yi Cui}
\email{cuiyi@ruc.edu.cn}
\affiliation{School of Physics and Beijing Key Laboratory of Opto{-}electronic Functional Materials \& Micro{-}nano Devices, Renmin University of China, 100872 Beijing, China.}
\affiliation{Key Laboratory of Quantum State Construction and Manipulation (Ministry of Education), Renmin University of China, 100872 Beijing, China.}
\author{Yuesheng Li}
\email{yuesheng\_li@hust.edu.cn}
\affiliation{Wuhan National High Magnetic Field Center and School of Physics, Huazhong University of Science and Technology, 430074 Wuhan, China}


\begin{abstract}
Frustrated magnets host strong quantum fluctuations that can suppress conventional magnetic order and give rise to exotic quantum phases such as spin liquids. In some cases, however, quantum fluctuations lift classical degeneracies and stabilize ordered states via an order-by-quantum-disorder mechanism. The spin-7/2 honeycomb antiferromagnet GdZnPO has recently been proposed as a spiral spin-liquid candidate arising from cooperative fluctuations among a subextensively degenerate manifold of spiral states. Here, we investigate the local magnetization and spin dynamics in GdZnPO using nuclear magnetic resonance. In an intermediate field regime between $\sim$3 T and the saturation field ($\sim$12 T), we observe a spatially uniform magnetization and persistent low-energy spin dynamics down to 0.033 K, with no detectable symmetry breaking, providing spectroscopic evidence for a spin-liquid state. At lower fields below $\sim$3 T, a weak stripe order emerges below $\sim$0.25 K; however, strong fluctuations persist, as indicated by a nearly temperature-independent and unusually large spin-lattice relaxation rate in the low-temperature limit. Our results demonstrate that spin-7/2 quantization weakly lifts the spiral degeneracy, stabilizing subtle magnetic order while preserving robust dynamics and spin-liquid phenomenology. These findings establish GdZnPO as a promising platform for exploring spin liquids in high-spin frustrated magnets down to the lowest accessible temperatures.
\end{abstract}


\maketitle

\section{Introduction}
Frustrated magnets can host exotic phases, such as spin liquids, characterized by fractionalized excitations, topological order, and long-range entanglement~\cite{balents2010spin,zhou2017quantum}. These phases hold promise for applications in topological quantum computation~\cite{nayak2008non,yao2013topologically}, spintronics~\cite{jungwirth2016antiferromagnetic,gao2020fractional,PhysRevLett.133.236704,zhao2025itinerant}, and adiabatic demagnetization refrigeration~\cite{zhao2025giant,PhysRevB.67.104421,xiang2024giant}, and may offer insights into the problem of high-temperature superconductivity~\cite{anderson1987resonating,RevModPhys.78.17}. In classical frustrated systems, accidental degeneracies can emerge, as exemplified by spiral spin liquids (SSLs). An SSL is characterized by a sub-extensive degeneracy of spiral configurations, where ground-state wave vectors form a continuous contour or surface in reciprocal space known as the spiral manifold~\cite{PhysRevResearch.4.023175}. At low temperatures, fluctuations drive the system across this manifold, suppressing conventional magnetic order.

In real materials, however, magnetic ions carry quantized spins and therefore inevitably experience quantum fluctuations. On the one hand, these fluctuations can suppress conventional magnetic order and stabilize quantum spin-liquid states at ultra-low temperatures, even down to 0 K; on the other hand, they may favor collinear or valence-bond correlations over classical spiral configurations, thereby lifting the classical degeneracy through an order-by-quantum-disorder mechanism~\cite{PhysRevLett.109.167201}. Within an SSL, ``order by disorder" refers to the mechanism by which quantum or thermal fluctuations lift the continuous ground-state degeneracy, thereby stabilizing magnetic order as the temperature is reduced~\cite{PhysRevB.81.214419,okumura2010novel}.

The $J_1$-$J_2$ frustrated antiferromagnet on the honeycomb lattice provides an effective prototypical framework for exploring quantum fluctuation effects  ~\cite{PhysRevB.81.214419,okumura2010novel,PhysRevB.100.224404,PhysRevB.106.035113,PhysRevResearch.4.013121,Bishop2012The,
PhysRevB.86.144404,PhysRevLett.110.127203,PhysRevB.92.195110},
\begin{multline}
\mathcal{H} = J_1\sum_{\langle j0,j1\rangle}\mathbf{S}_{j0}\cdot\mathbf{S}_{j1}+J_2\sum_{\langle\langle j0,j2\rangle\rangle}\mathbf{S}_{j0}\cdot\mathbf{S}_{j2}+D\sum_{j0}(S_{j0}^z)^2\\
-\mu_0Hg\mu_\mathrm{B}\sum_{j0}S_{j0}^z.
\label{eq1}
\end{multline}
Here, $J_1$ and $J_2$ represent the first ($\langle\rangle$) and second ($\langle\langle\rangle\rangle$) nearest-neighbor couplings, respectively, $D$ denotes the single-ion anisotropy relevant for spins $S$ $>$ 1/2, $g$ is the Land\'{e} factor, and the magnetic field $H$ is applied perpendicular to the honeycomb plane to preserve the $U$(1) symmetry. In the classical limit ($S$ $\to$ $\infty$), the SSL persists down to very low temperatures without order by thermal disorder, provided that $J_2$/$\mid$$J_1$$\mid$ $>$ 1/6, $D$ $\geq$ 0, and $H$ $<$ $H^*$~\cite{okumura2010novel,PhysRevResearch.4.013121,PhysRevLett.133.236704}. Here, $H^*$ marks the crossover/saturation field between the SSL and the field-polarized ferromagnetic state. In contrast, in the quantum limit ($S$ = 1/2 or 1),~Eq.~(\ref{eq1}) yields collinear magnetic orders, plaquette resonating valence-bond states, or deconfined quantum critical points separating them~\cite{Bishop2012The,PhysRevB.86.144404,PhysRevLett.110.127203,PhysRevB.92.195110}. Notably, no quantum analogue of the SSL has been identified for 0 $\leq$ $J_2$/$J_1$ $\leq$ 1, $D$ = 0, and $H$ = 0~\cite{Bishop2012The,PhysRevB.92.195110}. This model thus provides an ideal platform for exploring the stability of the SSL against quantum fluctuations at larger spin values.

Recently, the honeycomb antiferromagnet GdZnPO has emerged as a promising realization of the model in~Eq.~(\ref{eq1}), with spin quantum number $S$ = 7/2, $J_1$ $\sim$ $-$0.39 K, $J_2$ $\sim$ 0.57 K, $D$ $\sim$ 0.30 K, and $g$ = 2.01(2)~\cite{PhysRevLett.133.236704}. For $S$ $\gg$ 1/2, $J_2$/$\mid$$J_1$$\mid$ $\sim$ 1.5 ($>$ 1/2), and $D$ $>$ 0, the system at fields below $\mu_0H^*$ = $S$[2$D$+3$J_1$+9$J_2$+$J_1^2$/(4$J_2$)]/($g\mu_\mathrm{B}$) ($\sim$12 T), is expected to reside deeply within the SSL regime. The ground-state wave vectors \textbf{Q}$_\mathrm{G}$ form a ``spiral contour"---a continuous, closed curve surrounding the K point ($\{$1/3, 1/3$\}$) in reciprocal space---in contrast to conventional orders where \textbf{Q}$_\mathrm{G}$ is restricted to discrete values. This degeneracy yields residual low-temperature spin degrees of freedom, producing a finite residual specific heat $C_0$, analogous to that of an ideal gas. Experimentally, GdZnPO exhibits an unusually large magnetic specific heat persisting down to $\sim$50 mK ($\sim$0.4\%$\mid$$\Theta_\mathrm{cw}$$\mid$), consistent with the distinctive SSL behavior~\cite{yao2021generic}, $C_\mathrm{m}$ $\sim$ $C_0$+$C_1T$~\cite{PhysRevLett.133.236704,zhao2025itinerant}. Here, $C_0$ and $C_1$ are constant, and $\Theta_\mathrm{cw}$ = $-S$($S$+1)($J_1$+2$J_2$) ($\sim$$-$12 K) is the Curie-Weiss temperature~\cite{PhysRevLett.133.236704}. Owing to the low-energy structure of the spiral contour, GdZnPO single crystals at ultra-low temperatures simultaneously exhibit a large intrinsic magnetic thermal conductivity ($\kappa_{xx}^\mathrm{m}$ $\sim$ $\kappa_0$+$\kappa_1T$, where $\kappa_0$ $\sim$ 0.25 WK$^{-1}$m$^{-1}$ and $\kappa_1$ $\sim$ 0.97 WK$^{-2}$m$^{-1}$ are constant), a positive thermal Hall effect, and a giant magnetocaloric effect near $H^*$ (magnetic Gr\"{u}neisen ratio $\Gamma_\mathrm{m}$ $\sim$ 4.44 T$^{-1}$), surpassing the $\kappa_1$ and $\Gamma_\mathrm{m}$ values of other known magnetic insulators~\cite{zhao2025itinerant,zhao2025giant}. These exotic properties open the door to potential applications, such as adiabatic demagnetization refrigeration~\cite{PhysRevB.67.104421,xiang2024giant} and antiferromagnetic spintronics without magnetic-field leakage~\cite{jungwirth2016antiferromagnetic,gao2020fractional}. Despite its exotic bulk properties, direct spectroscopic and microscopic evidence of the SSL in GdZnPO has yet to be obtained.

\begin{figure*}
    \centering
	\includegraphics[width=0.8\textwidth]{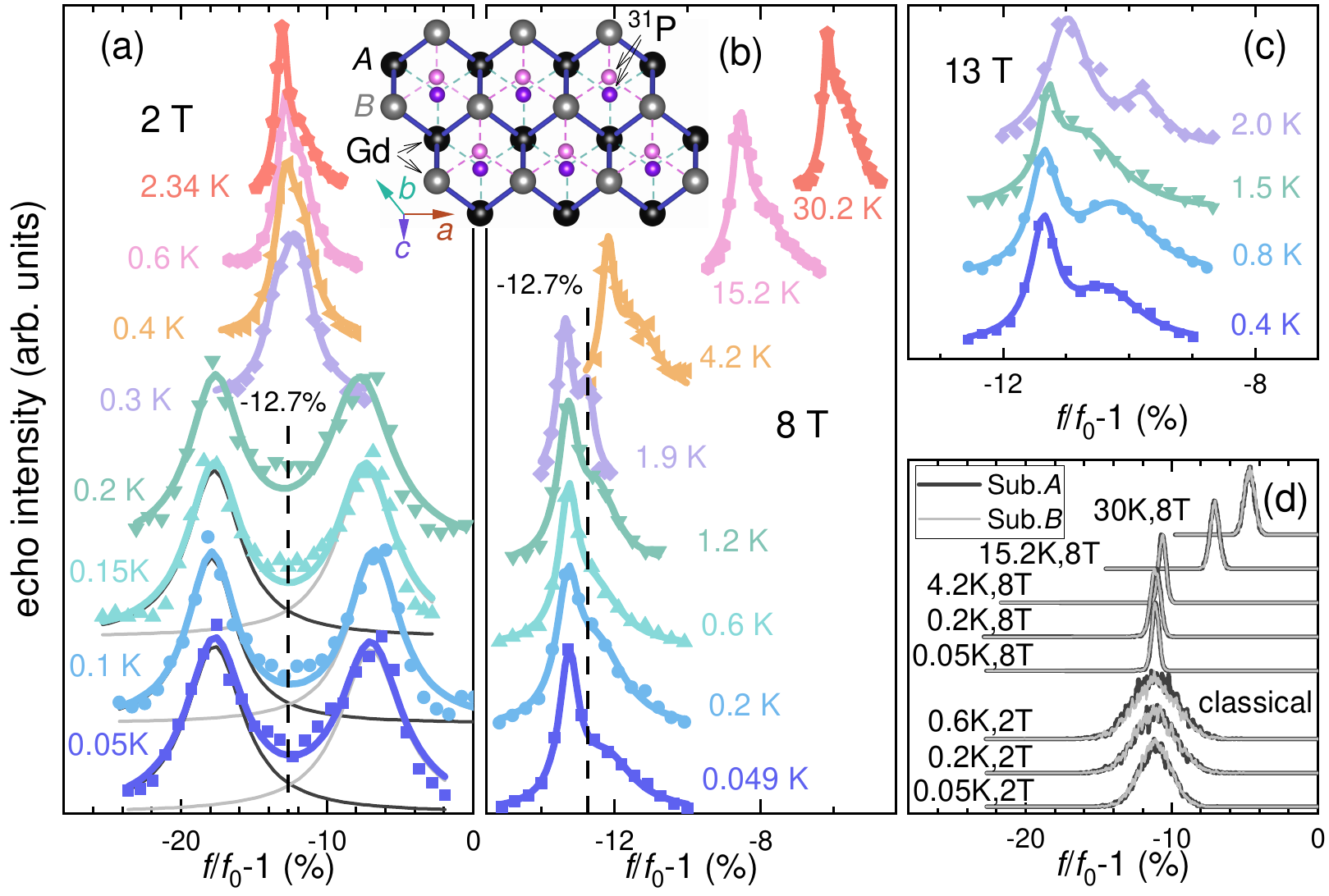}
	\caption{NMR spectra of GdZnPO. (a)-(c) Frequency-swept $^{31}$P NMR spectra measured at selected temperatures under 2, 8, and 13 T, respectively. The inset between (a) and (b) shows the GdZnPO crystal structure (Zn and O omitted for clarity). Each $^{31}$P nucleus probes its three nearest-neighbor Gd$^{3+}$ spins on either triangular sublattice \emph{A} or \emph{B}. In (a), black and gray lines denote the two Lorentzian components fitted below $\sim$0.25 K. (d) Simulated $^{31}$P NMR spectra from sublattices \emph{A} and \emph{B} at the classical level. Data are offset vertically for clarity, and the magnetic field is applied along the $c$ axis.}
	\label{fig1}
\end{figure*}

Among magnetic ions, Gd$^{3+}$ possesses the largest spin quantum number ($S$ = 7/2), and bulk measurements on GdZnPO generally support the classical SSL scenario at low temperatures and fields~\cite{PhysRevLett.133.236704,zhao2025itinerant,zhao2025giant}. Nevertheless, $S$ = 7/2 remains finite and inevitably introduces quantum fluctuations. Between $\sim$50 mK ($\sim$0.4\%$\mid$$\Theta_\mathrm{cw}$$\mid$) and $\sim$20 K ($\sim$2$\mid$$\Theta_\mathrm{cw}$$\mid$), the magnetic entropy increase reaches the saturation value, $\sim$3.00(3)$R$ln2 $\sim$ $R$ln(2$S$+1) at 0 T---substantially smaller than the classical Monte Carlo (MC) estimate of $\sim$7.5$R$ln2~\cite{PhysRevLett.133.236704}. Correspondingly, the residual specific heat observed in GdZnPO, $C_0$ = 0.9-1.4 JK$^{-1}$mol$^{-1}$, is much lower than the classical value $\sim$$R$/2 $\sim$ 4.2 JK$^{-1}$mol$^{-1}$~\cite{zhao2025itinerant}.

Our previous bulk measurements raise a key question: what is the precise role of quantum fluctuations in the SSL physics? To address this, a local probe is essential for resolving the microscopic symmetry and spin dynamics. While neutron scattering could, in principle, provide such information, it is severely hindered by the strong neutron absorption of Gd. In this work, we employ nuclear magnetic resonance (NMR) to investigate the local magnetization and spin dynamics in GdZnPO. At magnetic fields $H_\mathrm{c}$ $<$ $H$ $<$ $H^*$ ($\mu_0H_\mathrm{c}$ $\sim$ 3 T), the spin system shows no detectable symmetry breaking, maintaining spatially uniform magnetization and robust dynamics down to 33 mK ($\sim$0.3\%$\mid$$\Theta_\mathrm{cw}$$\mid$). In contrast, for $H$ $<$ $H_\mathrm{c}$, a spontaneous symmetry breaking between two sublattices emerges below $T_\mathrm{c}$ $\sim$ 0.25 K, while the spin dynamics remain strong to the lowest temperatures. These observations indicate that quantum fluctuations associated with the largest but finite $S$ = 7/2 favor a weak stripe-collinear order, slightly lifting the spiral degeneracy. Overall, our results demonstrate that GdZnPO retains well-defined SSL behavior despite the inevitable effects of spin quantization and quantum fluctuations.

\section{Methods}

\subsection{Sample preparation and NMR methods}
High-quality single crystals of GdZnPO were grown using the flux method (see Appendix~\ref{AppendixA})~\cite{PhysRevLett.133.236704}. $^{31}$P NMR measurements were carried out in a $^3$He-$^4$He dilution refrigerator down to 0.033 K. The NMR spectra were obtained using the standard spin-echo technique with the magnetic field applied along the crystallographic $c$ axis. The spin-lattice relaxation time $T_1$ was measured by the inversion-recovery method, and the nuclear magnetization was fitted using a stretched-exponential function appropriate for $I$ = 1/2 nuclei (see Appendix~\ref{AppendixA}).

\subsection{Bulk measurements and simulations}
Magnetization and specific-heat measurements were performed in magnetic fields up to 12 T applied along the $c$ axis, over 0.05-1.8 K using a $^3$He-$^4$He dilution refrigerator~\cite{PhysRevLett.133.236704,zhao2025itinerant,zhao2025giant}. Additional magnetization measurements up to 14 T were conducted between 1.9 and 150 K using a Physical Property Measurement System (Quantum Design). Standard classical MC simulations of the NMR spectra and specific heat were carried out on a $2\times72^2$ cluster with periodic boundary conditions, based on the spin Hamiltonian of GdZnPO (see Appendix~\ref{AppendixB})~\cite{PhysRevLett.133.236704}. Each simulation comprised 4000 MC steps at 200 temperature points, annealed gradually from 50 K to 0.05 K, with 2000 steps reserved for thermalization. Finite-size density-matrix renormalization group (DMRG) simulations were performed to study the ground-state properties of the GdZnPO spin Hamiltonian at zero field, employing $U$(1) symmetry (see Appendix~\ref{AppendixC}).

\section{Results}
\subsection{NMR spectra}

We performed $^{31}$P (natural abundance: 100\%, nuclear spin $I$ = 1/2, Zeeman factor $^{31}\gamma_\mathrm{n}$ = 17.235 MHz$\cdot$T$^{-1}$) NMR measurements on two independent single-crystal samples of GdZnPO (Appendix~\ref{AppendixA}). In GdZnPO, each $^{31}$P nucleus couples to three nearest-neighbor Gd$^{3+}$ ($S$ = 7/2) spins located on either the triangular sublattice $A$ or $B$~\cite{nientiedt1998equiatomic,lincke2008magnetic,PhysRevLett.133.236704}. The nearest-neighbor P-Gd distance ($\mid$P-Gd$\mid$ $\sim$ 3.00 \AA) closely matches the sum of the ionic radii of P$^{3-}$ (2.12 \AA) and Gd$^{3+}$ (0.94 \AA), confirming that $^{31}$P predominantly probes these three local moments. The nearest-neighbor distance to Gd$^{3+}$ spins on the opposite sublattice ($\mid$P-Gd$\mid'$ $\sim$ 5.34 \AA) is substantially larger, underscoring the highly local nature of the $^{31}$P NMR probe.

Figure~\ref{fig1} shows the NMR spectra of the GdZnPO single crystal (sample \#1) with the magnetic field applied perpendicular to the honeycomb plane, i.e., along the $c$ axis. Despite the high crystalline quality (see Appendix~\ref{AppendixA}), the NMR line shape is characteristic and nearly independent of temperature ($T$) and magnetic field ($H$) for $T$ $\geq$ 0.3 K or $\mu_0H$ $\geq$ 4 T [Figs.~\ref{fig1}(a)-\ref{fig1}(c)], where the system symmetry remains intact. At 30.2 K ($\sim$2.5$\mid$$\Theta_\mathrm{cw}$$\mid$), the system lies deep in the paramagnetic regime and retains full lattice symmetry [Fig.~\ref{fig1}(b)]. Even in the polarized ferromagnetic phase at 13 T ($>$$\mu_0H^*$), the overall line profile changes little [Fig.~\ref{fig1}(c)]. This robustness indicates that the observed characteristic line broadening is not intrinsic to the GdZnPO spin system, but most likely arises from internal strain, grain boundaries, or other crystalline imperfections that slightly distribute the hyperfine coupling $A_\mathrm{hf}$ between the $^{31}$P nuclei and the Gd$^{3+}$ spins ($|\Delta$$A_\mathrm{hf}$/$A_\mathrm{hf}|$ $\lesssim$ 2.5\% for sample \#1).

We also performed NMR measurements on sample \#2, which consisted of eight co-aligned smaller GdZnPO single crystals. The NMR line of \#2 is notably broader and exhibits a more pronounced characteristic profile (Appendix~\ref{AppendixA}), confirming the extrinsic origin of the fine structures of the NMR line (of \#1). When the intrinsic NMR line width is small (Appendix~\ref{AppendixA}), the characteristic spectral pattern arising from the distribution of $A_\mathrm{hf}$ becomes visible. Despite this sample-dependent broadening, the central resonance positions---i.e., the NMR shifts---are nearly identical between \#1 and \#2 (see Appendix~\ref{AppendixA}). Furthermore, no significant sample dependence is observed in the nuclear spin-lattice relaxation behavior (see Appendix~\ref{AppendixA}). The sample dependence of both the NMR shift and the spin-lattice relaxation rate is comparable to the experimental noise level and remains within the uncertainty of independent measurements. At $\sim$2 K and $<$12 T, the NMR shifts for both \#1 and \#2 samples approach a constant $K$ $\sim$ $-12.7$\%, matching the expected susceptibility for the honeycomb-lattice SSL ansatz (see below). These results demonstrate that our findings---mainly derived from the NMR shifts and nuclear spin-lattice relaxation rates---are intrinsic and robust against sample-specific variations.

\begin{figure*}
    \centering
	\includegraphics[width=0.78\textwidth]{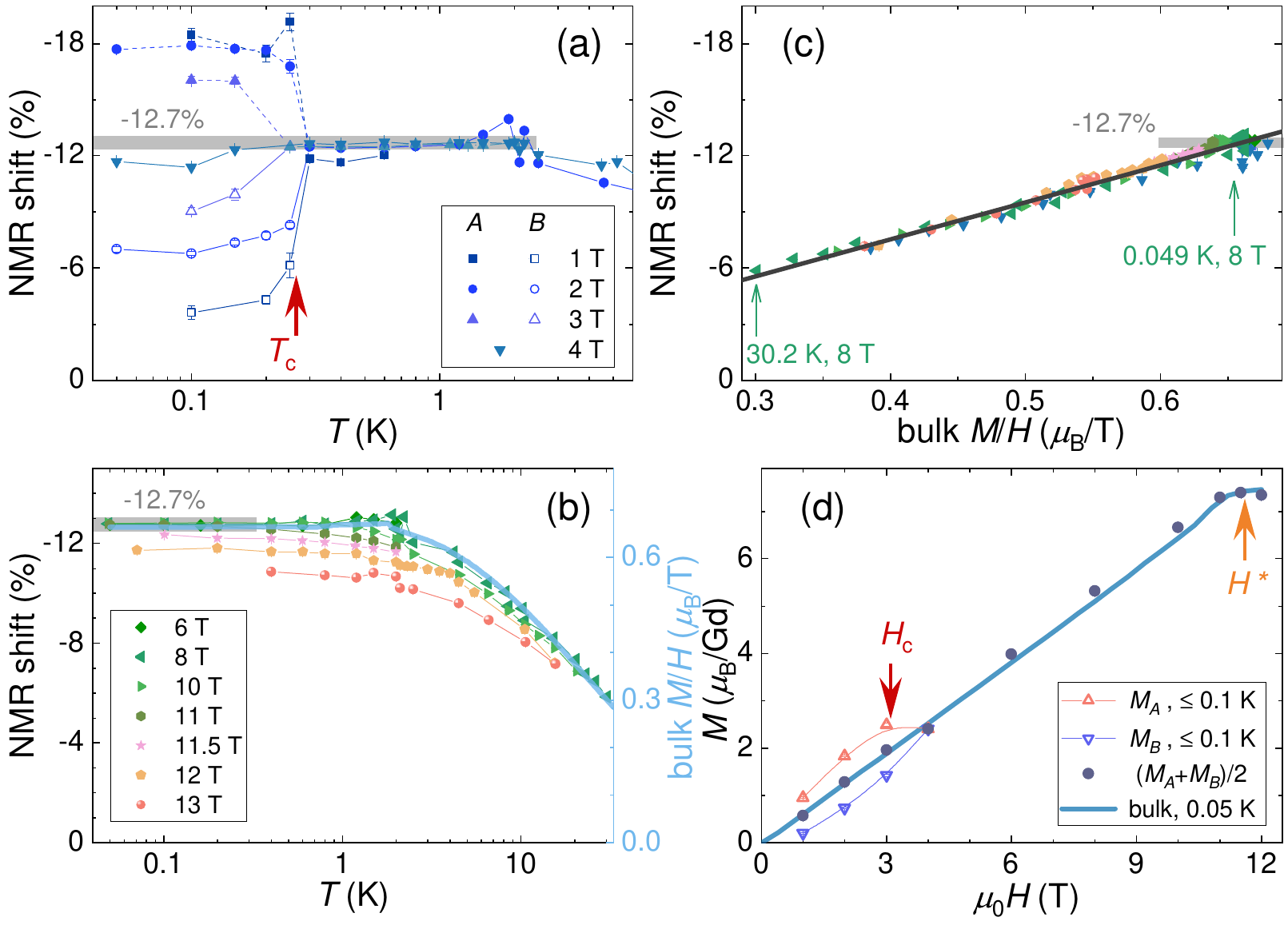}
	\caption{NMR shift and magnetization of GdZnPO. (a), (b) Temperature dependence of the NMR shift $K$($T$) measured under selected magnetic fields $H$. In (a), the crossover temperature $T_\mathrm{c}$ $\sim$ 0.25 K is indicated for $\mu_0H$ $\leq$ 3 T. In (b), bulk susceptibility $M$/$H$ is shown for $\mu_0H$ = 8 T. (c) Correlation between NMR shift $K$($T$,$H$) and bulk susceptibility $M$/$H$($T$,$H$) across the measurement range (0.049-30.2 K, 3-13 T), excluding the region $T$ $<$ $T_\mathrm{c}$ and $H$ $<$ $H_\mathrm{c}$ where $K$ splits [see (a)]. The black line represents a linear fit $K$ = $K_0$+$A_\mathrm{hf}M$/$H$, yielding $K_0$ = 0.004(2) and $A_\mathrm{hf}$ = $-$0.198(4) T/$\mu_\mathrm{B}$. (d) Field dependence of the NMR shift magnetization ($K$$-$$K_0$)$H$/$A_\mathrm{hf}$, averaged over $T$ $\leq$ 0.1 K, compared with the bulk magnetization at 0.05 K. Crossover fields $\mu_0H_\mathrm{c}$ $\sim$ 3 T and $\mu_0H^*$ $\sim$ 12 T are indicated. Magnetic field applied along the $c$ axis.}
	\label{fig2}
\end{figure*}

In contrast, at $T$ $<$ $T_\mathrm{c}$ $\sim$ 0.25 K and $\mu_0H$ $<$ $\mu_0H_\mathrm{c}$ $\sim$ 3 T, the NMR spectra exhibit a distinct line splitting, signaling an intrinsic spontaneous symmetry breaking [Fig.~\ref{fig1}(a)]. At the lowest temperatures, the spectrum resolves into two nearly symmetric and broadened Lorentzian peaks, and the fine structures evident at higher temperatures are smeared out. Owing to the symmetry breaking, the normalized full width at half maximum increases markedly to $\Delta$$f$/$f_0$ $\sim$ 5\% at $T$ $<$ $T_\mathrm{c}$ and $H$ $<$ $H_\mathrm{c}$, compared with only $\sim$1\% at higher temperatures or higher fields (see Appendix~\ref{AppendixA}). This intrinsic spectral splitting in GdZnPO contrasts sharply with classical ($S$ $\to$ $\infty$) simulations of Eq.~(\ref{eq1}), as illustrated in Fig.~\ref{fig1}(d).

Between $H_\mathrm{c}$ and $H^*$, the GdZnPO spin system remains partially polarized [Fig.~\ref{fig2}(d)] and is expected, within the classical model, to reside in the SSL regime at low temperatures. As shown in Fig.~\ref{fig1}(b), the NMR spectra exhibit neither discernible splitting nor broadening from 30.2 K down to 49 mK ($\sim$0.4\%$\mid$$\Theta_\mathrm{cw}$$\mid$), in excellent agreement with classical simulations [Fig.~\ref{fig1}(d)]. Furthermore, the fine structures of the NMR line remain nearly unchanged down to 49 mK for fields above $H_\mathrm{c}$ [Figs.~\ref{fig1}(b) and~\ref{fig1}(c)]. By contrast, in other spin-liquid candidates with inherent atomic mixing, such as YCu$_3$(OH)$_{6.5}$Br$_{2.5}$~\cite{Lu2022TheObservation,PhysRevB.109.104403} and ZnCu$_3$(OH)$_6$Cl$_2$~\cite{Evidence2015Fu,Khuntia2020Gapless}, the NMR line width typically broadens strongly at low temperatures due to spatially inhomogeneous magnetization arising from the structural disorder~\cite{Lu2022TheObservation,Chen2025The,Li2019YbMgGaO4}. The absence of such broadening is consistent with the lack of atomic mixing in GdZnPO~\cite{nientiedt1998equiatomic,lincke2008magnetic,PhysRevLett.133.236704}.

\subsection{NMR shift}

For $T$ $>$ $T_\mathrm{c}$ or $H$ $>$ $H_\mathrm{c}$, the NMR shift $K$ is obtained as $K$ = $\sum_jI_jf_j$/($f_0\sum_jI_j$)$-1$, where $I_j$ and $f_j$ denote the echo intensity and frequency of the $j$th data point, and $f_0$ = $^{31}\gamma_\mathrm{n}$$\mu_0H$ is the reference frequency. For $T$ $<$ $T_\mathrm{c}$ and $H$ $<$ $H_\mathrm{c}$, $K$ is determined from the fitted centers of the Lorentzian peaks [Fig.~\ref{fig1}(a)]. Below $T_\mathrm{c}$ $\sim$ 0.25 K and $\mu_0H_\mathrm{c}$ $\sim$ 3 T, the NMR shift exhibits a clear splitting, indicating a spontaneous symmetry breaking between the \emph{A} and \emph{B} sublattices [Fig.~\ref{fig2}(a)]. Around the crossover temperature $T^*$ $\sim$ 2 K, classical MC simulations predict a peak or kink in the out-of-plane susceptibility~\cite{PhysRevLett.133.236704}, consistent with spontaneous chiral-symmetry breaking for $J_2$/$\mid$$J_1$$\mid$ $\sim$ 1.5 $>$ 0.7~\cite{PhysRevResearch.4.013121}. However, both the experimental NMR shift and the bulk susceptibility show no discernible anomalies near $T^*$ [see Figs.~\ref{fig2}(a) and~\ref{fig2}(b)], apart from noise arising from the measurement-mode transition between the $^4$He cryostat and the $^3$He-$^4$He refrigerator around 2 K. Moreover, the NMR spectra exhibit neither a noticeable change in line shape (Fig.~\ref{fig1}) nor any splitting (Fig.~\ref{fig2}) near $T^*$ across 1-13 T, excluding conventional magnetic order.

\begin{figure*}
	\centering
	\includegraphics[width=0.96\textwidth]{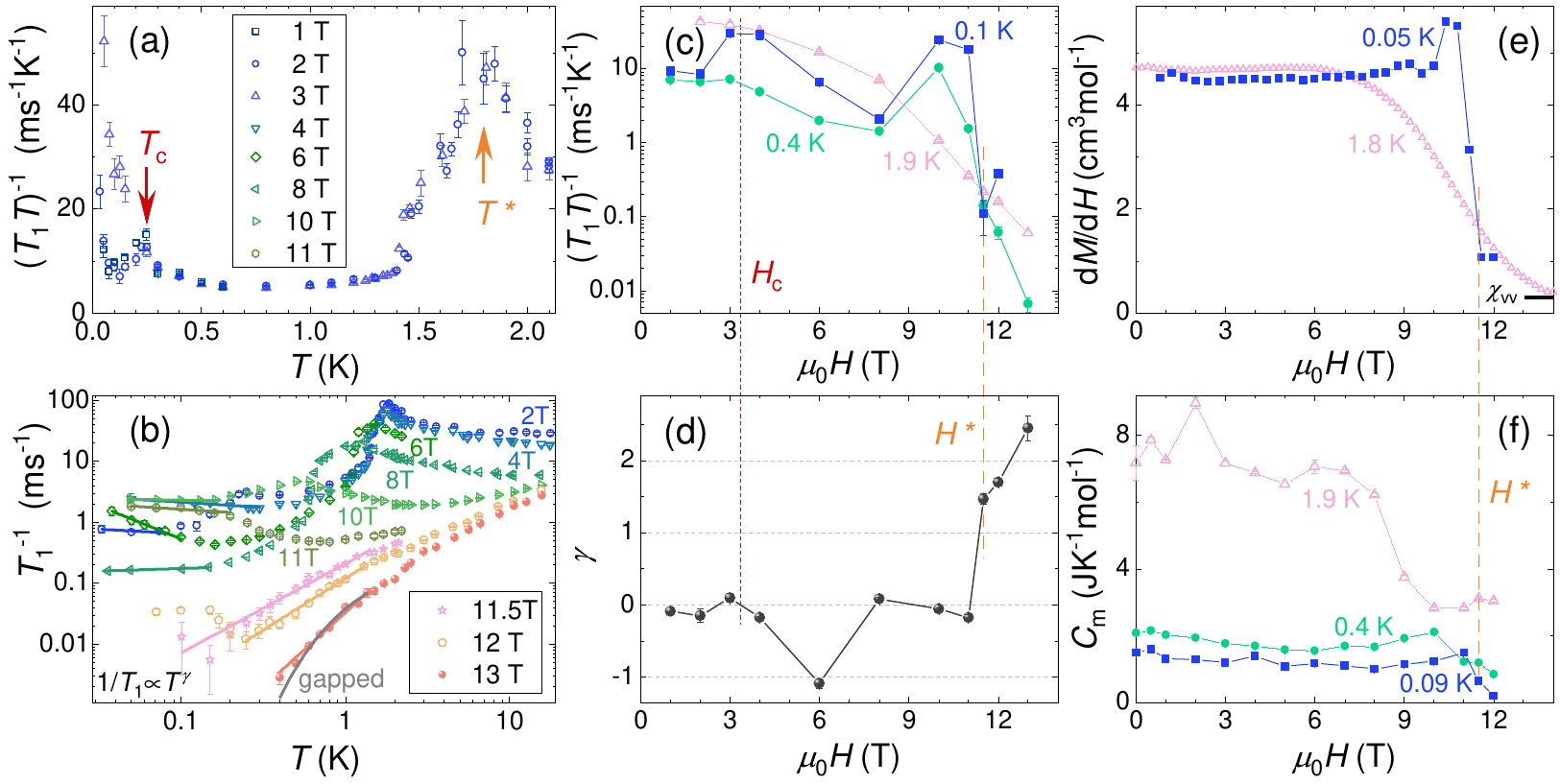}
	\caption{Nuclear spin-lattice relaxation rate of GdZnPO. (a), (b) Temperature dependence of the $^{31}$P nuclear spin-lattice relaxation rate 1/$T_1$ measured under selected magnetic fields $H$. In (a), crossover temperatures $T_\mathrm{c}$ $\sim$ 0.25 K and $T^*$ $\sim$ 2 K are indicated for $\mu_0H$ $\leq$ 2 T. In (b), colored lines represent power-law fits 1/$T_1$ $\propto$ $T^\gamma$, with the fitted exponent $\gamma$ shown in (d). The gray line corresponds to a fit to the 13 T data using 1/$T_1$ $\propto$ $\exp$($-\Delta$/$T$), yielding a gap $\Delta$ = 2.24(7) K. (c), (d) Field dependence of 1/($T_1T$) at selected temperatures and the exponent $\gamma$. Crossover fields $\mu_0H_\mathrm{c}$ $\sim$ 3 T and $\mu_0H^*$ $\sim$ 12 T are indicated. (e), (f) Field dependence of bulk susceptibility (d$M$/d$H$) and magnetic specific heat ($C_\mathrm{m}$) at selected temperatures. The Van Vleck susceptibility $\chi_\mathrm{vv}$ $\sim$ 0.3 cm$^3$mol$^{-1}$ is indicated in (e)~\cite{PhysRevLett.133.236704}. Magnetic field applied along the $c$ axis.}
	\label{fig3}
\end{figure*}

Above $T_\mathrm{c}$ or $H_\mathrm{c}$, the NMR shift $K$ scales nearly linearly with the bulk susceptibility $M$/$H$ at each temperature and field [Fig.~\ref{fig2}(b)], consistent with a spatially uniform magnetization and the SSL picture in GdZnPO. The bulk $M$/$H$ increases monotonically with decreasing temperature, and the $K$-$M$/$H$ data at various temperatures and fields collapse onto a common scaling relation, $K$ $\sim$ $K_0$+$A_\mathrm{hf}$$M$/$H$, with a temperature- and field-independent chemical shift $K_0$ = 0.004(2) ($\approx$0) and a hyperfine coupling constant $A_\mathrm{hf}$ = $-$0.198(4) T/$\mu_\mathrm{B}$ [Fig.~\ref{fig2}(c)]. At low temperatures and $H$ $<$ $H^*$,  $K$ $\sim$ $-$12.7\% (for $H$ $>$ $H_\mathrm{c}$ or $T$ $>$ $T_\mathrm{c}$), or the average of the two split values (for $H$ $<$ $H_\mathrm{c}$ and $T$ $<$ $T_\mathrm{c}$), corresponding to a susceptibility ($M$/$H$)$_K$ = ($K$$-$$K_0$)/$A_\mathrm{hf}$ $\sim$ 4.6 cm$^3$mol$^{-1}$, is nearly independent of $T$ and $H$. This agrees well with the bulk susceptibility [Fig.~\ref{fig3}(e)]~\cite{zhao2025giant}, and the classical SSL ansatz $\chi_\mathrm{class}$ = $\mu_0N_\mathrm{A}g^2\mu_\mathrm{B}^2/[2D$+$3J_1$+$9J_2$+$J_1^2/(4J_2)]$+$\chi_\mathrm{vv}$ $\sim$ 4.4 cm$^3$mol$^{-1}$~\cite{zhao2025itinerant,zhao2025giant}, where $\chi_\mathrm{vv}$ $\sim$ 0.3 cm$^3$mol$^{-1}$ is the Van Vleck contribution~\cite{PhysRevLett.133.236704}.

At temperatures below $\sim$0.1 K, the NMR shift splitting exhibits near saturation with respect to temperature for $H$ $<$ $H_\mathrm{c}$ [Figs.~\ref{fig1}(a) and~\ref{fig2}(a)]. The NMR-derived magnetization, ($K$$-$$K_0$)$H$/$A_\mathrm{hf}$, averaged over data at $T$ $\leq$ 0.1 K, is shown in Fig.~\ref{fig2}(d). For $H$ $<$ $H_\mathrm{c}$, the magnetization splits into $M_A$ and $M_B$ ($M_A$ $\neq$ $M_B$) due to spontaneous symmetry breaking between sublattices \emph{A} and \emph{B}, whereas for $H$ $>$ $H_\mathrm{c}$ it becomes uniform ($M_A$ = $M_B$). The average ($M_A$+$M_B$)/2 agrees well with the low-$T$ bulk magnetization, validating the scaling relation. Along the $c$ axis, the extracted ground-state order parameters, $\mid$$M_A$$-$$M_B$$\mid$/$M_\mathrm{s}$ = 0.107(5), 0.157(4), and 0.152(9) at 1, 2, and 3 T, respectively ($M_\mathrm{s}$ = $gS$ = 7 $\mu_\mathrm{B}$/Gd), are far smaller than 2, indicating a weak antiferromagnetic order below $T_\mathrm{c}$ and $H_\mathrm{c}$ in GdZnPO.

\subsection{Spin dynamics}

The nuclear spin-lattice relaxation of GdZnPO is well described by a stretched exponential function over the entire measured $T$-$H$ range (Appendix~\ref{AppendixA}),
\begin{equation}
M_\mathrm{n}(t) = M_\mathrm{n}(\infty)[1-\alpha~\mathrm{e}^{-(\frac{t}{T_1})^\beta}],
\label{eq2}
\end{equation}
where $M_\mathrm{n}(\infty)$ and $\alpha$ (1 $<$ $\alpha$ $<$ 2) are scale factors. The stretching exponent $\beta$ remains close to 1 (with $\mid$$\beta$$-$$1$$\mid$ within the fitting uncertainty) above $\sim$ 1.5 K, and slightly decreases to 0.65-1.0 at lower temperatures for $\mu_0H$ $<$ 13 T (see Appendix~\ref{AppendixA}). In contrast, the relaxation rate 1/$T_1$ shows a pronounced dependence on temperature and field (Fig.~\ref{fig3}), but no evident sample dependence (Appendix~\ref{AppendixA}).

1/($T_1T$) probes spin fluctuations perpendicular to the applied magnetic field, i.e., within the honeycomb plane, following 1/($T_1T$) $\sim$ $\sum_\mathbf{q}\chi''(\mathbf{q},f_0)$/($hf_0$)~\cite{Moriya1963The,Bloembergen1948Relaxation,Redfield1957On}. 
For $\mu_0H$ $\leq$ 4 T, 1/($T_1T$) shows a broad maximum around $T^*$ $\sim$ 2 K [Fig.~\ref{fig3}(a)], closely resembling the behavior of the magnetic specific heat~\cite{PhysRevLett.133.236704,zhao2025itinerant}. This indicates a gradual crossover from the high-temperature paramagnetic phase to the putative SSL with the spin structure factor concentrating along the spiral contour~\cite{PhysRevLett.133.236704}, rather than the sharp transition expected classically from chiral symmetry breaking at $J_2$/$\mid$$J_1$$\mid$ $\sim$ 1.5 $>$ 0.7~\cite{PhysRevResearch.4.013121}. With increasing magnetic field, the crossover temperature $T^*$ shifts progressively lower [Fig.~\ref{fig3}(b)], consistent with the specific-heat results~\cite{PhysRevLett.133.236704,zhao2025itinerant}.

\begin{table}[h!]
\caption{Low-temperature spin-lattice relaxation rate (1/$T_1$) behavior for ordered magnetic insulators (top) and gapless spin systems (bottom). The exponent $\gamma$ is defined by 1/$T_1$ $\propto T^{\gamma}$. Abbreviations: KAF (kagome antiferromagnet), TAF (triangular antiferromagnet), TFIC (transverse-field Ising chain), HKAF (hyperkagome antiferromagnet), HAF (honeycomb antiferromagnet), and EDT-BCO ([EDT-TTF-CONH$_2$]$^+_2$[BABCO$^-$]).}
\begin{center}
\begin{tabular}{c c c c}
\multicolumn{4}{c}{Ordered magnetic insulators} \\
\hline
\hline
compound & $\gamma$ & conditions ($T$, $H$)\\
\hline
SrMn$_2$P$_2$~\cite{Sangeetha2021First} & 1.5 & $<$40 K\\
\hline
Nd$_{0.85}$Sr$_{0.15}$NiO$_2$~\cite{Cui2021NMR} & 2 & $<$40 K, 9 T\\
\hline
NiGa$_2$S$_4$~\cite{PhysRevB.77.054429} & 3 & $<$0.5 K &\\
\hline
NH$_4$Fe(PO$_3$F)$_2$~\cite{PhysRevB.111.184435} & 3 & $<$6 K, 1.3 T\\
\hline
EuO~\cite{PhysRevB.72.014428} & 7/2 & $>$14 K, 0 T\\
\hline
CaMn$_2$P$_2$~\cite{Sangeetha2021First} & 5 & 20-70 K\\
\hline
K$_2$V$_3$O$_8$~\cite{PhysRevB.100.054406} & 5 & $<$4 K, 3 and 6.6 T\\
\hline
Na$_2$Co$_2$TeO$_6$~\cite{PhysRevB.106.224416} & 5 & $<$13 K, 1-9 T\\
\hline
Na$_2$Cu$_3$O(SO$_4$)$_3$~\cite{PhysRevB.107.245134} & 5.37 & $<$3 K, 5.6 and 7.3 T\\
 & gapped & 8.6-33 T\\
\hline
Na$_2$BaCo(PO$_4$)$_2$~\cite{gsk8-1k9q} & gapped & 0.7 and 1 T\\
\hline
\hline
\end{tabular}
\end{center}

\begin{center}
\begin{tabular}{c c c c}
\multicolumn{4}{c}{Gapless spin systems with strong low-energy fluctuations} \\
\hline
\hline
compound & $\gamma$ & system & conditions\\
\hline
Li$_9$Fe$_3$(P$_2$O$_7$)$_3$(PO$_4$)$_2$ & $-$1 & $S$ = 5/2 & 1.5-5 K\\
~\cite{PhysRevLett.127.157202} & & KAF & $\leq$6 T\\
\hline
CoNb$_2$O$_6$ & $-$0.75 & $S$ = 1/2 & $<$7 K\\
~\cite{PhysRevX.4.031008} & & TFIC & $\sim$critical field\\
\hline
Cu$_2$(C$_5$H$_{12}$N$_2$)$_2$Cl$_4$ & $-$0.5 & $S$ = 1/2 & \\
~\cite{Chaboussant1998Nuclear} & & spin ladder & $\sim$critical field \\
\hline
EDT-BCO~\cite{Szirmai2020Quantum} & $-$0.8 & $S$ = 1/2 TAF & $<$10 K\\
\hline
YCu$_3$(OH)$_{6.5}$Br$_{2.5}$ & $-$0.2 & $S$ = 1/2 & $<$5.1 K\\
~\cite{Lu2022TheObservation,PhysRevB.109.104403} & & KAF &\\
\hline
NaYbSe$_2$~\cite{Zhu2023Fluctuating} & 0 & $S$ = 1/2 TAF & $<$2 K, 4.5 T\\
\hline
6HB-Ba$_3$NiSb$_2$O$_9$~\cite{PhysRevB.93.214432} & 0 & $S$ = 1 TAF & $<$200 K, 12 T\\
\hline
PbCuTe$_2$O$_6$~\cite{PhysRevLett.116.107203} & 0.4 & $S$ = 1/2 HKAF & 2-20 K, 7 T\\
\hline
ZnCu$_3$(OH)$_6$Cl$_2$~\cite{PhysRevLett.100.087202} & 0.7 & $S$ = 1/2 KAF & $<$20 K\\
\hline
ZnCu$_3$(OH)$_6$SO$_4$~\cite{Gomilsek2017Field} & 0.8 & $S$ = 1/2 KAF & $<$0.2 K, 4.7 T\\
\hline
(H,Li)$_6$Ru$_2$O$_6$~\cite{PhysRevB.110.L241102} & 0.8-0.9 & $S$ = 1/2 Kitaev & $<$10 K, 1.6, 2.4 T\\
\hline
GdZnPO (this work) & $-1$-0 & $S$ = 7/2 HAF & $<$0.2 K, $<$12 T\\
\hline
\hline
\end{tabular}
\end{center}
\label{tab1}
\end{table}

Under 1 or 2 T, a small peak appears in 1/($T_1T$) at $T_\mathrm{c}$ $\sim$ 0.25 K [Fig.~\ref{fig3}(a)], consistent with the NMR line splitting [Figs.~\ref{fig1}(a) and \ref{fig2}(a)] and indicating the emergence of weak order below $T_\mathrm{c}$ and $H_\mathrm{c}$ in GdZnPO. Remarkably, in the low-temperature limit, 1/($T_1T$) shows a pronounced upturn as $T$ decreases to 33 mK ($\sim$ 0.3\%$\mid$$\Theta_\mathrm{cw}$$\mid$), following 1/($T_1T$) $\propto$ $T^{\gamma-1}$ with $-$1 $\lesssim$ $\gamma$ $\lesssim$ 0 for fields below $\mu_0H^*$ $\sim$ 12 T (Fig.~\ref{fig3}). This anomalous behavior contrasts sharply with ordered magnetic insulators, where 1/($T_1T$) $\propto$ $T^{\gamma-1}$ with $\gamma$ $>$ 1 or 1/$T_1$ $\propto$ e$^{-\Delta/T}$ with a finite gap $\Delta$ (see Table~\ref{tab1}). In contrast, the persistent upturn of 1/($T_1T$) in GdZnPO resembles that found in gapless spin systems with strong low-energy fluctuations. Representative examples include the classical spin-liquid candidate Li$_9$Fe$_3$(P$_2$O$_7$)$_3$(PO$_4$)$_2$, the transverse-field Ising chain system CoNb$_2$O$_6$ near the critical field, the spin ladder Cu$_2$(C$_5$H$_{12}$N$_2$)$_2$Cl$_4$ near the critical field, and many quantum spin-liquid candidates (see Table~\ref{tab1}). The power-law behavior of 1/$T_1$ $\propto$ $T^{\gamma}$ with $-$1 $\lesssim$ $\gamma$ $\lesssim$ 0 thus evidences robust, gapless spin dynamics persisting as $T$ $\to$ 0 K for $H$ $<$ $H^*$ [see Figs.~\ref{fig3}(a)-\ref{fig3}(d)], likely originating from cooperative fluctuations among subextensively degenerate spiral states in the putative SSL of GdZnPO.

For $H$ $<$ $H^*$, both $-$log$_{10}$($T_1T$) and $\gamma$ are nearly field-independent at low temperatures in GdZnPO [Figs.~\ref{fig3}(c) and~\ref{fig3}(d)], consistent with the SSL ansatz and its field-independent spiral contour (see Appendix~\ref{AppendixB}). Above $H^*$, $-$log$_{10}$($T_1T$) drops abruptly, and $\gamma$ increases sharply above 1, indicating a crossover from the dynamic SSL regime to a conventional polarized ferromagnetic phase. Remarkably, at 0.4 K, 1/$T_1$ exceeds 1000 s$^{-1}$ in the putative spin-liquid state (below $H^*$)---an unusually large value that is roughly three orders of magnitude greater than the 2.6 s$^{-1}$ measured in the conventional polarized phase at 13 T [above $H^*$, see Fig.~\ref{fig3}(b)]. The field dependence of $-$log$_{10}$($T_1T$) closely mirrors previously reported behaviors of the bulk susceptibility (d$M$/d$H$) and magnetic specific heat ($C_\mathrm{m}$) in GdZnPO [Figs.~\ref{fig3}(e) and~\ref{fig3}(f)]~\cite{zhao2025itinerant,zhao2025giant}. Below $T_\mathrm{c}$ $\sim$ 0.25 K, $-$log$_{10}$($T_1T$) is only slightly suppressed for $H$ $<$ $H_\mathrm{c}$ [Fig.~\ref{fig3}(c)], corresponding to the formation of weak magnetic order. As $\mu_0H$ approaches $\mu_0H^*$ $\sim$ 12 T, $-$log$_{10}$($T_1T$), d$M$/d$H$, and $C_\mathrm{m}$ exhibit a local maximum, followed by a sharp decrease, at the lowest temperatures (Fig.~\ref{fig3}), possibly due to residual quantum fluctuations associated with the finite spin $S$ = 7/2. At 13 T, the low-temperature 1/$T_1$ data can alternatively be described by a gapped function, 1/$T_1$ $\propto$ e$^{-\Delta/T}$, with $\Delta$ = 2.24(7) K [Fig.~\ref{fig3}(b)], in reasonable agreement with spin-wave calculations for GdZnPO~\cite{zhao2025itinerant}. At 14 T and 1.9 K, d$M$/d$H$ approaches $\chi_\mathrm{vv}$ $\sim$ 0.3 cm$^3$mol$^{-1}$~\cite{PhysRevLett.133.236704}, indicating near-complete polarization of the spin system.

\section{Phase diagram and discussion}

\begin{figure*}
	\centering
	\includegraphics[width=0.85\textwidth]{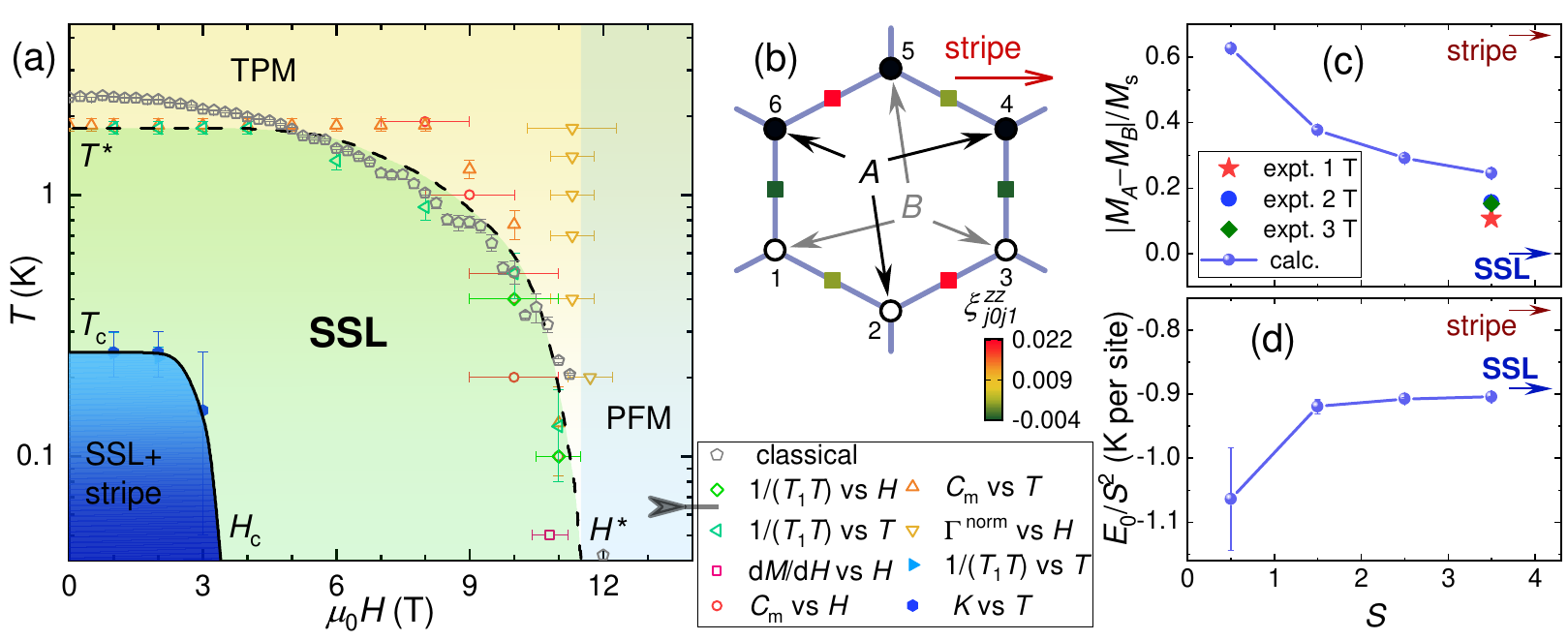}
	\caption{Phase diagram of GdZnPO. (a) Magnetic field-temperature ($H$-$T$) phase diagram extracted from NMR shift ($K$), spin-lattice relaxation rate (1/$T_1$), bulk susceptibility (d$M$/d$H$), magnetic specific heat ($C_\mathrm{m}$)~\cite{zhao2025itinerant}, and normalized Gr\"{u}neisen ratio $\Gamma^\mathrm{norm}$ = ($H$/$T$)(d$T$/d$H$)~\cite{zhao2025giant}. 
The thermal paramagnetic (TPM) phase occurs where thermal energy ($k_\mathrm{B}T$) dominates spin-spin exchange interactions, whereas the polarized ferromagnetic (PFM) phase is the low-temperature state where an external magnetic field overcomes these interactions to align the spins uniformly. (b) Ground-state ($\mid$GS$\rangle$) correlations of the central hexagon, $\xi_{j0j1}^{zz}$ = $\langle$GS$\mid$$S_{j0}^zS_{j1}^z$$\mid$GS$\rangle$/$S^2$, computed for an $S$ = 7/2 72-site cluster (Appendix~\ref{AppendixC}). Solid and hollow circles denote opposite spin orientations. (c), (d) Spin quantum number ($S$) dependence of $\mid$$M_A$$-$$M_B$$\mid$/$M_\mathrm{s}$ = $\sqrt{\langle\mathrm{GS}\mid(\sum_{j0=1,3,5}S_{j0}^z-\sum_{j1=2,4,6}S_{j1}^z)^2\mid\mathrm{GS}\rangle}$/(3$S$) and ground-state energy per site ($E_0$), extrapolated from finite-size simulations at 0 T (Appendix~\ref{AppendixC}). Arrows indicate the classical ($S$ $\to$ $\infty$) values of the pure spiral spin liquid (SSL) and stripe state. Experimental values at 1, 2, and 3 T are shown for comparison in (c). All simulations use the interaction parameters of GdZnPO~\cite{PhysRevLett.133.236704}, and $H$ $\parallel$ $c$ (or $z$).}
	\label{fig4}
\end{figure*}

The phase diagram of GdZnPO is summarized in Fig.~\ref{fig4}(a). Below $T_\mathrm{c}$ $\sim$ 0.25 K and $\mu_0H_\mathrm{c}$ $\sim$ 3 T, a spontaneous symmetry breaking between sublattices \emph{A} and \emph{B} is clearly evidenced by the symmetric splitting of the NMR spectrum [Fig.~\ref{fig1}(a)] and the corresponding shifts [Fig.~\ref{fig2}(a)]. This behavior sharply contrasts with the ground spiral states of the classical ($S$ $\to$ $\infty$) $J_1$-$J_2$ frustrated honeycomb model [Eq.~(\ref{eq1})]. From $S$ $\to$ $\infty$ to $S$ = 1/2, the ground state of~Eq.~(\ref{eq1}) is dramatically altered in the studied parameter regime~\cite{Bishop2012The,PhysRevB.86.144404,PhysRevLett.110.127203,PhysRevB.92.195110}, giving rise to order by quantum disorder. It is therefore natural to expect that a similar mechanism operates in the GdZnPO spin system. Using the experimentally determined exchange parameters, $J_1$ $\sim$ $-$0.39 K and $J_2$ $\sim$ 0.57 K~\cite{PhysRevLett.133.236704}, we find a clear zigzag-stripe antiferromagnetic order (Appendix~\ref{AppendixC}) in the quantum limit ($S$ = 1/2), with a substantial reduction of the ground-state energy relative to the classical spiral value $E_0$/$S^2$ = $-J_1^2$/(8$J_2$)$-$3$J_2$/2 $\sim$ $-$0.89 K per site [see Fig.~\ref{fig4}(d)]. The stripe order breaks the symmetry between the triangular sublattices \emph{A} and \emph{B}, producing up-up-down and up-down-down spin configurations, respectively (or vice versa), naturally accounting for the observed finite value of $\mid$$M_A$$-$$M_B$$\mid$/$M_\mathrm{s}$.

As $S$ increases and quantum fluctuations are reduced, the calculated $\mid$$M_A$$-$$M_B$$\mid$/$M_\mathrm{s}$ gradually decreases, and $E_0$/$S^2$ rises, approaching the classical SSL value of $\sim$$-$0.89 K per site [Figs.~\ref{fig4}(c) and~\ref{fig4}(d)]. For $S$ = 7/2 at 0 T, our quantum many-body simulations indicate a possible coexistence of spiral and zigzag-stripe correlations in the ground state, yielding a competitive ground-state energy $E_0$/$S^2$ = $-$0.9040(2) K per site, slightly lower than the classical SSL value. However, the calculated $\mid$$M_A$$-$$M_B$$\mid$/$M_\mathrm{s}$ = 0.246(14) for $S$ = 7/2 remains larger than the experimental value of $\leq$ 0.16 [see Fig.~\ref{fig4}(c)], likely reflecting limitations of traditional finite-size approaches in simulating quantum spin systems with incommensurate spiral order~\cite{PhysRevLett.133.176502}. Nevertheless, the quantum simulations semi-quantitatively capture the NMR observations at $T$ $<$ $T_\mathrm{c}$ and $H$ $<$ $H_\mathrm{c}$ in GdZnPO.

The applied field makes precise quantum many-body simulations challenging, and must disfavor the antiferromagnetic order [Fig.~\ref{fig4}(b)]. In contrast, the low-temperature SSL remains robust against external fields up to $\mu_0H^*$ $\sim$ 12 T. This robustness is well supported by the NMR shift [Figs.~\ref{fig2}(b) and~\ref{fig2}(d)], nuclear spin-lattice relaxation rate [Figs.~\ref{fig3}(b) and~\ref{fig3}(c)], and bulk measurements [Figs.~\ref{fig3}(e) and~\ref{fig3}(f)], naturally explaining why the pure SSL is stabilized in the range $H_\mathrm{c}$ $<$ $H$ $<$ $H^*$ without spontaneous symmetry breaking, down to at least 49 mK [Fig.~\ref{fig1}(b)]. By comparison, the stripe order in GdZnPO is weak: (1) the measured order parameter $\mid$$M_A$$-$$M_B$$\mid$/$M_\mathrm{s}$ $\leq$ 0.16 is much lower than the 2/3 expected for a pure stripe order; (2) only a weak peak appears in the relaxation rate around $T_\mathrm{c}$ at fields below 3 T [Fig.~\ref{fig3}(a)]; (3) $\mu_0H_\mathrm{c}$ ($\sim$3 T) $\ll$ $\mu_0H^*$ ($\sim$12 T) and $T_\mathrm{c}$ ($\sim$0.25 K) $\ll$ $T^*$ ($\sim$2 K); and (4) no clear anomaly is observed around $H_\mathrm{c}$ or $T_\mathrm{c}$ in bulk measurements~\cite{PhysRevLett.133.236704,zhao2025itinerant,zhao2025giant}. Taken together, these observations support the coexistence of a dominant SSL with weak stripe order at $T$ $<$ $T_\mathrm{c}$ and $H$ $<$ $H_\mathrm{c}$ in the phase diagram of GdZnPO [Fig.~\ref{fig4}(a)].

Realizing an SSL in a real material down to ultra-low temperatures of $\sim$0.4\%$\mid$$\Theta_\mathrm{cw}$$\mid$ is extremely challenging, as quantum fluctuations~\cite{PhysRevLett.109.167201,Bishop2012The,PhysRevB.92.195110}, perturbative interactions~\cite{gao2024}, and even thermal fluctuations~\cite{okumura2010novel} can lift the accidental degeneracy and induce magnetic order. Fortunately, GdZnPO benefits from the largest spin $S$ = 7/2 and negligible perturbative interactions beyond~Eq.~(\ref{eq1}), due to the highly localized nature of the 4$f$ electrons and the zero orbital angular momentum of Gd$^{3+}$ ions (Appendix~\ref{AppendixB})~\cite{PhysRevLett.133.236704}. Moreover, MC simulations demonstrate that the SSL remains stable down to at least $T$ $\sim$ 0.3\%$\mid$$\Theta_\mathrm{cw}$$\mid$ against thermal fluctuations, based on the classical version of the GdZnPO spin Hamiltonian~\cite{PhysRevResearch.4.013121,PhysRevLett.133.236704}. Despite the weak stripe order observed at $T$ $<$ $T_\mathrm{c}$ and $H$ $<$ $H_\mathrm{c}$, the magnetism of the $S$ = 7/2 honeycomb antiferromagnet GdZnPO can still be well approximated within the simplest classical SSL framework [Fig.~\ref{fig4}(a)].

Except for $\gamma$ $\sim$ $-$1 and a classical spin-liquid-like behavior~\cite{PhysRevLett.127.157202} at $\sim$6 T, we observe $\gamma$ $\sim$ 0 [Fig.~\ref{fig3}(d)] and an almost $T$-independent 1/$T_1$ [Fig.~\ref{fig3}(b)] as $T$ $\to$ 0 K across all other fields below $H^*$. This unexpectedly robust behavior resembles that of gapless quantum spin-liquid candidates~\cite{Zhu2023Fluctuating,PhysRevB.93.214432,Li2016Muon}, despite the high spin $S$ = 7/2 in GdZnPO. This nearly $T$-independent 1/$T_1$ suggests that quantum fluctuations associated with $S$ = 7/2 help stabilize a resilient spin-liquid ground state without evident residual magnetic entropy at $T$ $\sim$ 0.4\%$\mid$$\Theta_\mathrm{cw}$$\mid$~\cite{PhysRevLett.133.236704} against perturbations, making GdZnPO a promising new route toward quantum (spiral) spin liquids~\cite{niggemann2019classical}.

\section{CONCLUSIONS}

Our NMR results further confirm the dominance of SSL physics in GdZnPO down to 33 mK ($\sim$0.3\%$\mid$$\Theta_\mathrm{cw}$$\mid$) below $H^*$, as evidenced by the robust spin dynamics and the small order parameter $\mid$$M_A$$-$$M_B$$\mid$/$M_\mathrm{s}$ $\leq$ 0.16. Importantly, the observed spin-liquid behavior is intrinsic and not a mimicry induced by structural disorder: no atomic-mixing disorder has been reported, and no discernible broadening of the NMR line is detected from 30.2 K down to 49 mK (above $H_\mathrm{c}$). The spin quantization of $S$ = 7/2 in GdZnPO plays a double-edged role. On one hand, it induces a weak stripe collinear order via an order-by-quantum-disorder mechanism, embedded within the dominant SSL at $T$ $<$ $T_\mathrm{c}$ and $H$ $<$ $H_\mathrm{c}$. On the other hand, the residual quantum fluctuations associated with $S$ = 7/2 act to smear the transition and preserve the underlying symmetries, thereby stabilizing the dynamic SSL against perturbations down to ultra-low temperatures, at least in the regime $H_\mathrm{c}$ $<$ $H$ $<$ $H^*$. Our results thus shed new light on the search for spin-liquid candidates in high-spin frustrated magnets that preserve strong spin dynamics and fluctuations down to the lowest accessible temperatures.

\section*{Acknowledgments}

Y.L. gratefully acknowledges Shang Gao and Wei Li for helpful discussion in Ganzhou. This work was supported by the National Key R\&D Program of China (Grant No. 2023YFA1406500), the Strategic Priority Research Program of Chinese Academy of Sciences (No. XDB0500202), the Scientific Research Innovation Capability Support Project for Young Faculty (No. ZYGXQNJSKYCXNLZCXM-M26), the National Natural Science Foundation of China (No. 12374156, No. 12134020, No. 12274153, No. 12322403, and No. 12347107), the Youth Innovation Promotion Association of Chinese Academy of Sciences (No. 2021004), and the Fundamental Research Funds for the Central Universities (No. HUST: 2020kfyXJJS054).

\section*{Data availability}

The data are not publicly available. All raw data corresponding to the findings in this manuscript are available from the authors upon reasonable request.

\appendix

\section{Sample dependence, NMR line width, nuclear spin-lattice relaxation, and stretching exponent}\label{AppendixA}

\begin{figure}[h]
	\includegraphics[width=0.49\textwidth]{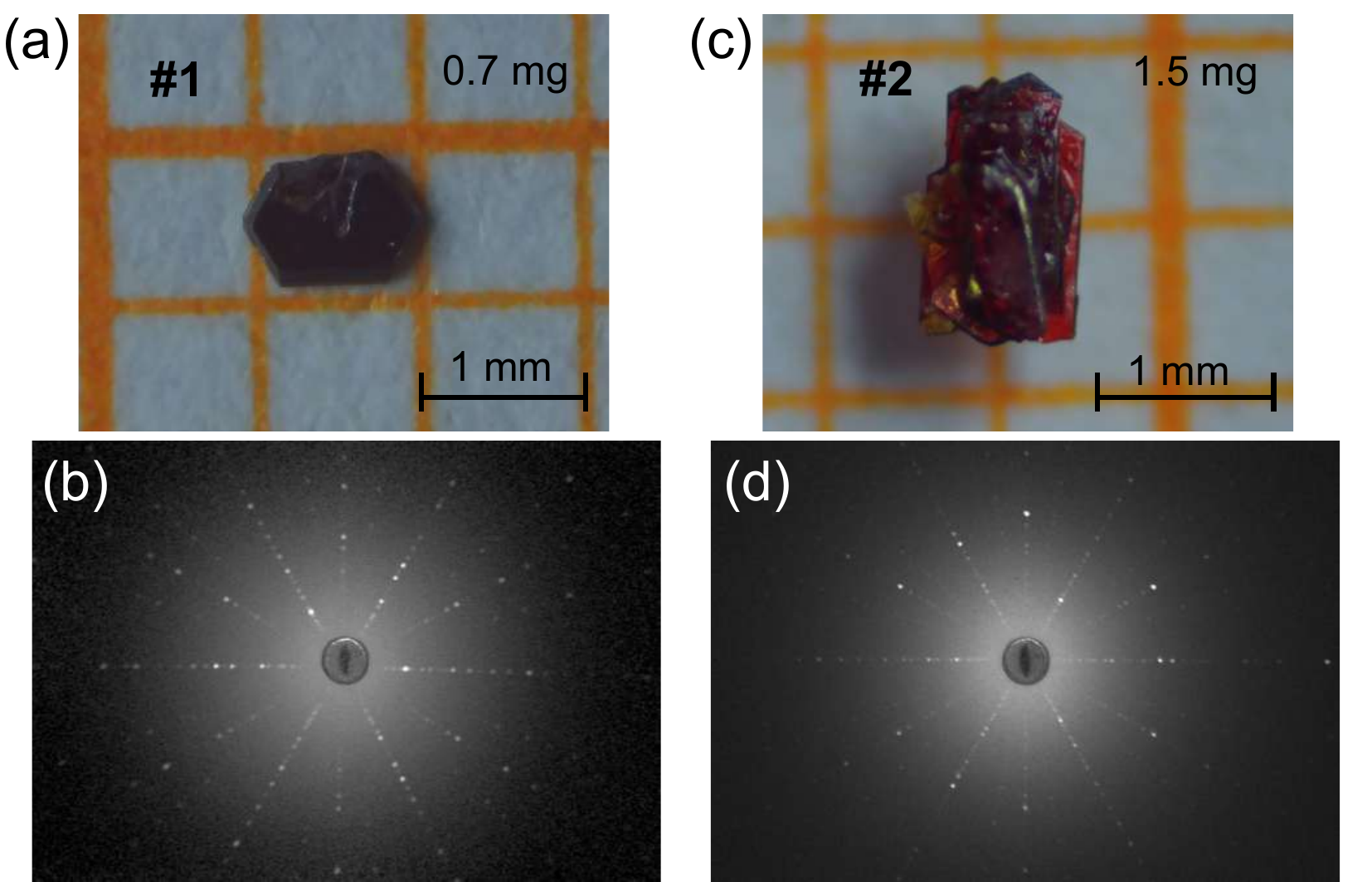}
	\caption{NMR samples \#1 and \#2. (a), (c) Single-crystal samples \#1 and \#2 of GdZnPO used for $^{31}$P NMR measurements. (b) Laue x-ray diffraction pattern obtained from the upper surface of sample \#1. For sample \#2, eight thin crystals were co-aligned along the $c$ axis using GE varnish. (d) Laue pattern measured on a representative single crystal from sample \#2.}
	\label{figs1}
\end{figure}

The as-grown GdZnPO crystals are reddish-brown platelets oriented along the $ab$ plane, as confirmed by Laue x-ray diffraction. Nine crystals exhibiting sharp and well-defined Laue patterns were selected for the $^{31}$P (abundance: 100\%, nuclear spin $I$ = 1/2) NMR investigation, as shown in Fig.~\ref{figs1}. The thickest and largest crystal, sample \#1 [mass $\sim$0.7 mg; Fig.~\ref{figs1}(a)], was comprehensively studied under magnetic fields between 1 and 13 T (generated by a superconducting magnet) applied along the crystallographic $c$ axis, over temperatures from $\sim$30 K down to 0.033 K, using either a $^4$He cryostat (2-30.2 K) or a $^3$He-$^4$He dilution refrigerator (0.033-2 K). The magnetic field was applied perpendicular to the honeycomb lattice to preserve the $U$(1) symmetry, ensuring that the GdZnPO spin system remains within the putative SSL regime, with the degenerate contour unchanged for $H$ $<$ $H^*$~\cite{PhysRevB.106.035113}. The crossover field is given by $\mu_0H^*$ = $S[2D$+$3J_1$+$9J_2$+$J_1^2/(4J_2)]/(g\mu_\mathrm{B})$ ($\sim$12 T), where $S$ = 7/2, $J_1$ $\sim$ $-$0.39 K, $J_2$ $\sim$ 0.57 K, $D$ $\sim$ 0.30 K, and $g$ = 2.01(2) (see below)~\cite{PhysRevLett.133.236704,zhao2025giant,zhao2025itinerant}.

\begin{figure}
	\includegraphics[width=0.49\textwidth]{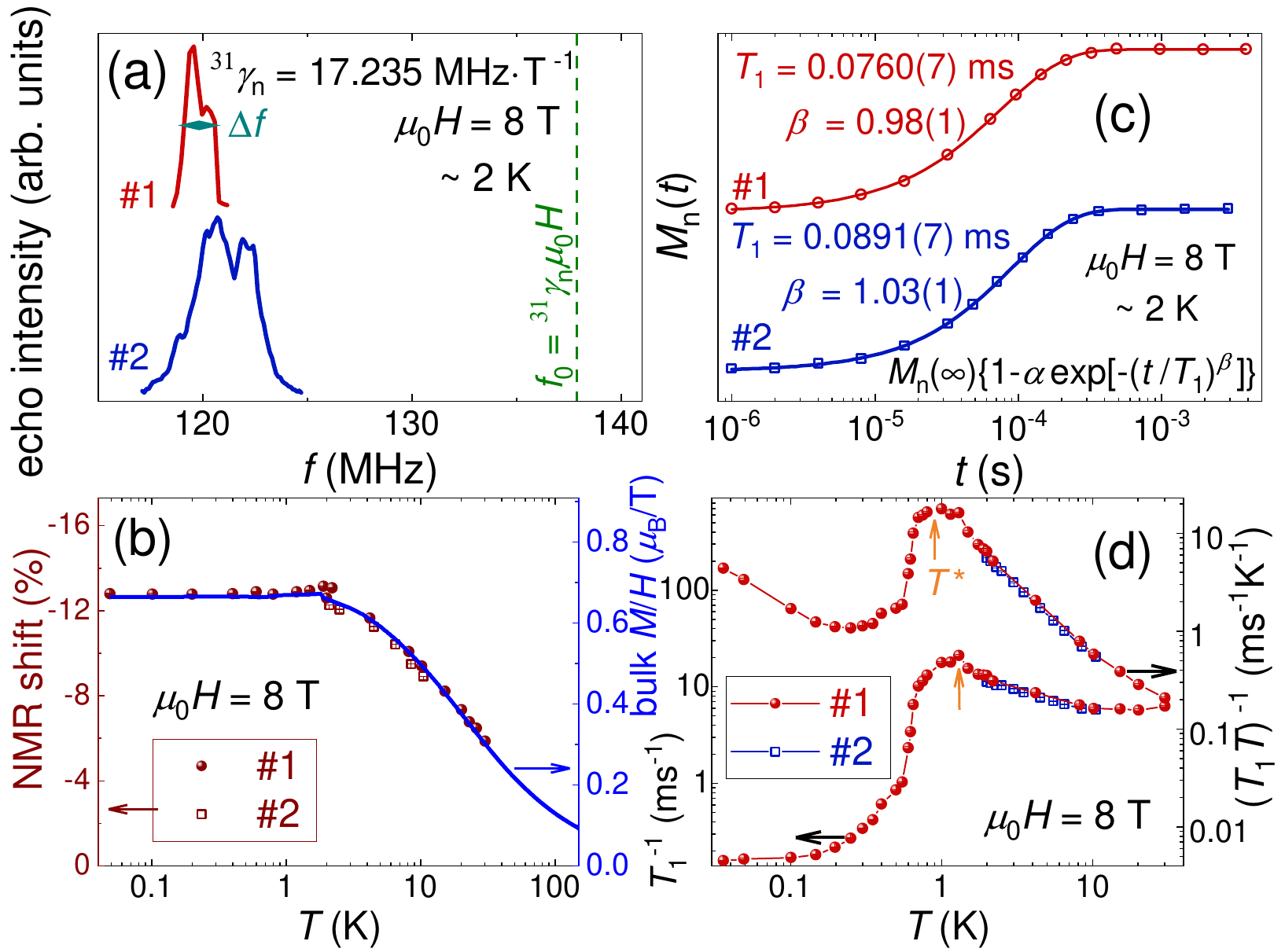}
	\caption{Sample dependence of NMR results on GdZnPO. (a) Frequency-swept $^{31}$P NMR spectra recorded at 8 T and $\sim$2 K for samples \#1 and \#2. The full width at half maximum ($\Delta f$) for \#1 is indicated. (b) Temperature dependence of the Knight shift for \#1 and \#2 compared with the bulk susceptibility $M$/$H$ at 8 T. (c) Nuclear spin-lattice relaxation curves measured at 8 T and $\sim$2 K for \#1 and \#2; solid lines represent fits using a stretched-exponential function. (d) Temperature dependence of the $^{31}$P spin-lattice relaxation rate (1/$T_1$) for \#1 and \#2. Vertical arrows indicate crossover temperatures $T^*$ $\sim$ 1.3 and 0.9 K, determined by 1/$T_1$ and 1/($T_1T$), respectively. Data in (a) and (c) are offset vertically for clarity.}
	\label{figs2}
\end{figure}

Despite the high crystalline quality (Fig.~\ref{figs1}), the $^{31}$P NMR spectra of sample \#1 display a sharp but characteristic multi-featured pattern, rather than a simple single Lorentzian peak [Fig.~\ref{figs2}(a)]. This pattern remains nearly unchanged across the entire measured range, except within the stripe-ordered regime ($T$ $<$ $T_\mathrm{c}$ $\sim$ 0.25 K and $\mu_0H$ $<$ $\mu_0H_\mathrm{c}$ $\sim$ 3 T), as discussed in the main text. At high temperatures (30.2 K $\gg$ $T^*$ $\sim$ 2 K), the system lies deep within the thermally paramagnetic phase, while at 13 T ($>$$\mu_0H^*$) and low temperatures, the spins are nearly fully polarized. Therefore, the detailed spectral features most likely originate from internal strain, grain boundaries, and/or other minor crystal imperfections that slightly distribute the hyperfine coupling $A_\mathrm{hf}$ between the $^{31}$P nuclei and the Gd$^{3+}$ spins ($|\Delta$$A_\mathrm{hf}$/$A_\mathrm{hf}|$ $\lesssim$ 2.5\% for sample \#1), since $A_\mathrm{hf}$ is highly sensitive to the P-Gd bonds. In GdZnPO single crystals, such structural imperfections can also induce noticeable sample dependence in the thermal transport properties by influencing the mean free paths of itinerant quasiparticles~\cite{zhao2025itinerant}.

When the intrinsic NMR line width of the GdZnPO spin system is small, the characteristic pattern of the spectrum becomes visible, and its sample dependence may also emerge. To verify this, we prepared an additional NMR sample, \#2, by co-aligning eight thinner and smaller crystals along the $c$ axis [Fig.~\ref{figs1}(c)]. As shown in Fig.~\ref{figs2}(a), sample \#2 exhibits evidently broader NMR lines with richer fine structures compared to the single crystal \#1. The broader and more structured spectra of \#2 can be attributed to the presence of multiple crystals and the consequently enhanced distribution of the hyperfine coupling $A_\mathrm{hf}$.

Despite the evident sample dependence of the NMR spectral pattern, the Knight shift shows negligible variation between samples \#1 and \#2 [Fig.~\ref{figs2}(b)]. Furthermore, the nuclear spin-lattice relaxation rate also displays negligible sample dependence [Figs.~\ref{figs2}(c) and ~\ref{figs2}(d)]. In this work, NMR measurements below $\sim$2 K were performed primarily on the single crystal \#1, which exhibits narrower NMR lines. Above $\sim$2 K, both the Knight shift and the relaxation rate 1/$T_1$ are nearly sample-independent; thus, the sample index (\#1 or \#2) is not distinguished for data presented in this temperature range [see Figs.~\ref{figs2}(b) and~\ref{figs2}(d), and the main text).

\begin{figure}
	\includegraphics[width=0.49\textwidth]{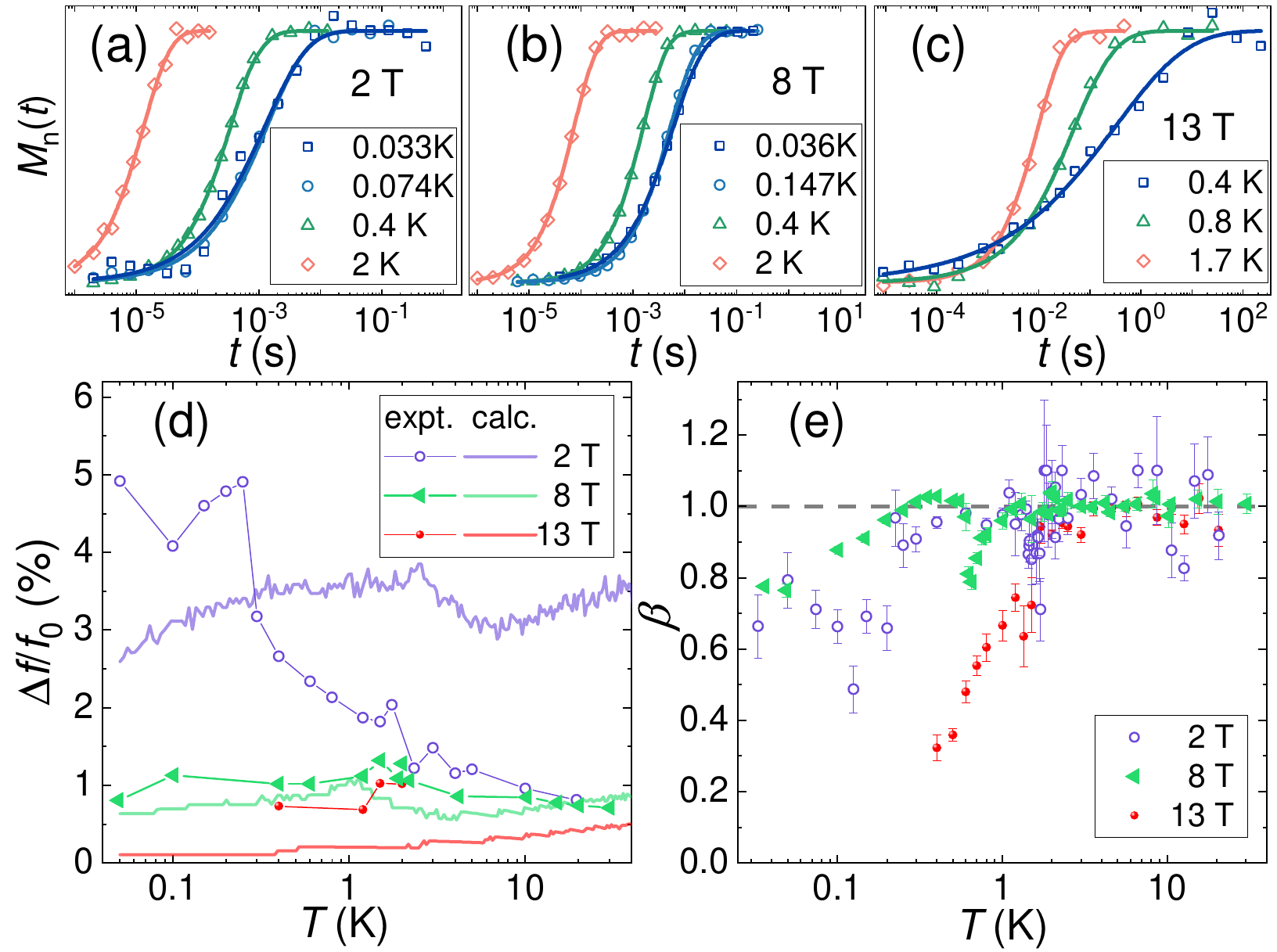}
	\caption{Nuclear spin-lattice relaxation, NMR line width, and stretching exponent of GdZnPO. (a)-(c) Nuclear spin-lattice relaxation curves measured at selected magnetic fields and temperatures. (d) Normalized full width at half maximum of the NMR line, $\Delta f$/$f_0$, measured on sample \#1, compared with classical ($S$ $\to$ $\infty$) simulations. Here $f_0$ = $^{31}\gamma_\mathrm{n}$$\mu_0H$, with $^{31}\gamma_\mathrm{n}$ = 17.235 MHz$\cdot$T$^{-1}$ as the Zeeman factor of $^{31}$P. (e) Temperature dependence of the stretching exponent $\beta$.}
	\label{figs3}
\end{figure}

In the stripe-ordered regime ($T$ $<$ $T_\mathrm{c}$ and $H$ $<$ $H_\mathrm{c}$), the normalized NMR line width is markedly enhanced to $\Delta f$/$f_0$ $\sim$ 5\% [Fig.~\ref{figs3}(d)], and the line shape becomes a simple Lorentzian without characteristic fine structures. In contrast, for $H$ $>$ $H_\mathrm{c}$, the NMR line width remains small and nearly temperature-independent, $\Delta f$/$f_0$ $\sim$ 1\%, despite the large temperature decrease from 30.2 K down to 0.049 K, in excellent agreement with classical MC simulations [Fig.~\ref{figs3}(d)]. This behavior provides strong evidence for the absence of appreciable interaction randomness in GdZnPO. By comparison, in other spin-liquid candidates with significant atomic mixing, such as YCu$_3$(OH)$_{6.5}$Br$_{2.5}$~\cite{Lu2022TheObservation,PhysRevB.109.104403} and ZnCu$_3$(OH)$_6$Cl$_2$~\cite{Evidence2015Fu,Khuntia2020Gapless}, the NMR line width typically broadens substantially at low temperatures, reflecting the spatially inhomogeneous magnetization induced by inherent interaction randomness~\cite{Lu2022TheObservation}.

Across the entire measured $T$-$H$ range, the nuclear spin-lattice relaxation can be well fitted by a stretched-exponential function for the central transition of $I$ = 1/2 nuclei [Eq.~(\ref{eq2})], as exemplified in Figs.~\ref{figs2}(c) and~\ref{figs3}(a)-\ref{figs3}(c). Above $\sim$ 1.5 K, the relaxation is essentially single-exponential, with $|$$\beta-1$$|$ comparable to the fitting uncertainty [Figs.~\ref{figs2}(c) and~\ref{figs3}(e)]. At lower temperatures, $\beta$ slightly decreases to 0.65-1.0 for $\mu_0H$ $<$ 13 T. In the low-temperature limit and at most fields below $\mu_0H^*$ $\sim$ 12 T, the raw relaxation curves become nearly temperature-independent [Figs.~\ref{figs3}(a) and~\ref{figs3}(b)], indicating that 1/$T_1$ is also nearly temperature-independent (see Fig.~\ref{fig3} in the main text).

\section{Spin Hamiltonian of GdZnPO and its classical spiral ground state}\label{AppendixB}

Under a magnetic field ($H$) applied along the $c$ axis, the $S$ = 7/2 Hamiltonian of GdZnPO can be approximated by Eq.~(\ref{eq1})~\cite{PhysRevLett.133.236704}. The $g$ factor was determined to be $g$ = 2.01(2), very close to the free-electron value $g_e$ = 2.00~\cite{PhysRevLett.133.236704}, consistent with the zero orbital angular momentum ($L$ = 0) of the Gd$^{3+}$ ground-state electronic configuration (4$f^7$, total angular momentum $J$ = $S$ = 7/2).

The single-ion anisotropy ($D$) is a common feature of spin systems with $S$ $>$ 1/2~\cite{Liu2021Frustrated}, arising from the crystalline electric-field (CEF) effect. In GdZnPO, the $D$ term accounts for the observed magnetization anisotropy at low temperatures ($\sim$1.9 K)~\cite{PhysRevLett.133.236704} and is primarily governed by the CEF Hamiltonian term $B_2^0O_2^0$. Here, $O_2^0$ = $3(S_{j0}^z)^2$$-$$S(S$+$1)$ is the Stevens operator, and the CEF parameter $B_2^0$ is evaluated within the point-charge model as~\cite{Bauer2009Magnetism}
\begin{equation}
B_2^0 \approx -\frac{\alpha_J\langle r^2\rangle e^2}{16\pi\epsilon_0}\sum_j\frac{Z_j(3\cos^2\theta_j-1)}{R_j^3},
\label{eqs3}
\end{equation}
where $\{R_j$, $\theta_j\}$ are the polar coordinates of the $j$th point charge relative to the Gd$^{3+}$ ion, $Z_j$ denotes the effective charge (in units of $e$), and $\langle r^2\rangle$ =  0.2428 \AA$^2$ is the radial matrix element for Gd$^{3+}$~\cite{Bauer2009Magnetism}. Based on the crystallographic structure of GdZnPO~\cite{PhysRevLett.133.236704}, we obtain $B_2^0$ $\sim$ $-$443.4$\alpha_J$ K. Using the experimental value $B_2^0$ = $D$/3 $\sim$ 0.10 K, we estimate the Stevens factor $\alpha_J$ $\sim$ $-$2$\times$10$^{-4}$, whose absolute value is about two orders of magnitude smaller than those of other rare-earth ions with ground-state $L$ $\neq$ 0 and $J$ $\neq$ 0~\cite{Bauer2009Magnetism}. Consequently, the single-ion anisotropy ($D$ $\sim$ 0.30 K) in GdZnPO can be naturally attributed to higher-order mixing with the excited electronic configurations of Gd$^{3+}$ having $L$ $\neq$ 0.

Alternatively, an easy-plane anisotropy of the superexchange interactions with an energy scale of $\sim$ $D$ $\sim$ 0.3 K could, in principle, account for the magnetization anisotropy observed at low temperatures. However, such a large anisotropy---comparable to the exchange strengths ($|J_1|$ or $J_2$)---is physically unrealistic and inconsistent with the small deviation of the $g$ factor from the free-electron value, $|g-g_e|$/$g_e$ $\lesssim$ 1\%~\cite{PhysRevLett.133.236704,PhysRev.120.91}. Therefore, the anisotropy of superexchange interactions in GdZnPO must be negligibly small.

\begin{figure}
	\includegraphics[width=0.49\textwidth]{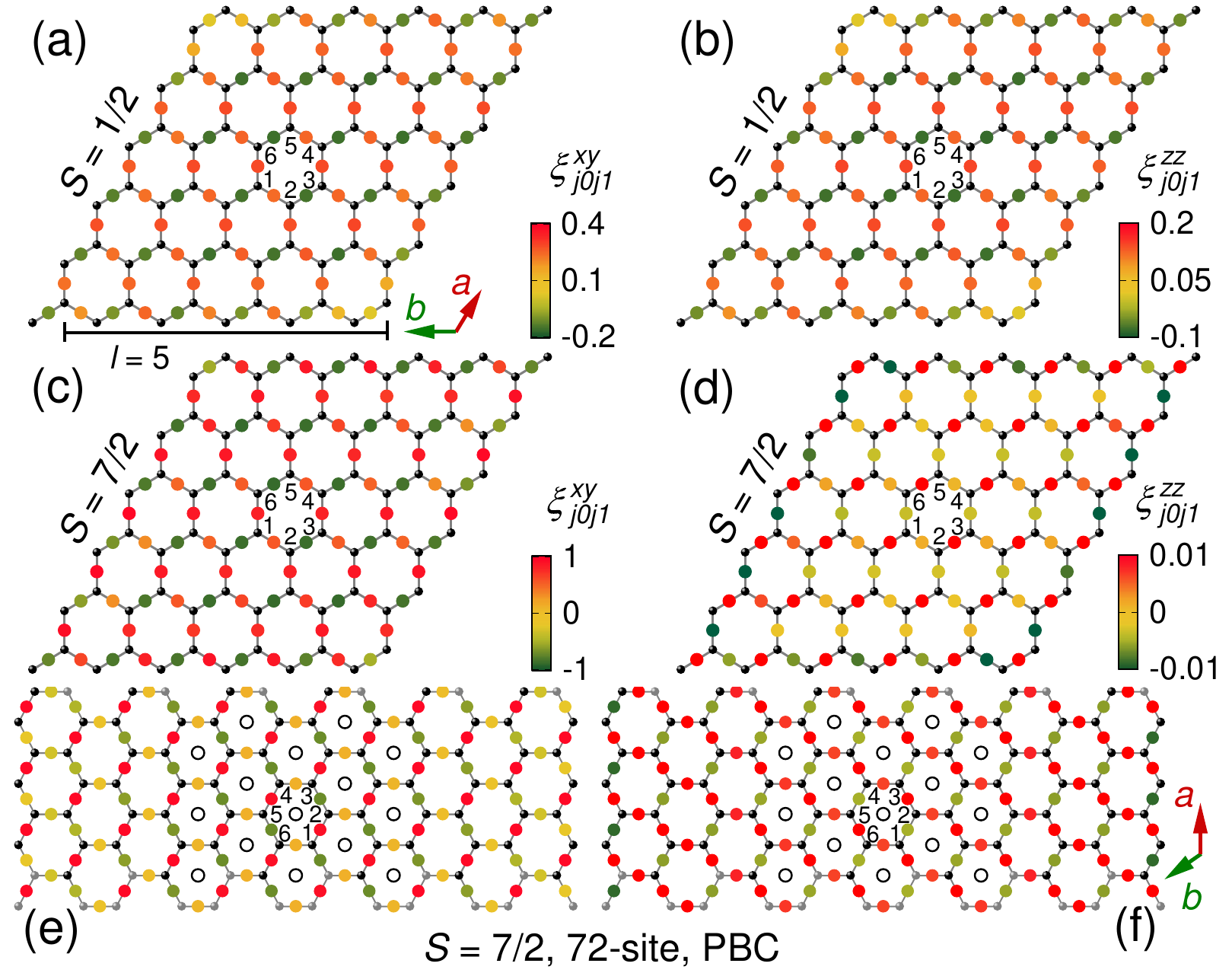}
	\caption{DMRG simulations of ground-state spin correlations in GdZnPO. (a), (b) Nearest-neighbor spin-spin correlations $\xi_{j0j1}^{xy}$ and $\xi_{j0j1}^{zz}$ calculated for an $S$ = 1/2 72-site cluster with open boundary conditions (OBC). (c), (d) Corresponding $\xi_{j0j1}^{xy}$ and $\xi_{j0j1}^{zz}$ for an $S$ = 7/2 72-site cluster with OBC. (e), (f) $\xi_{j0j1}^{xy}$ and $\xi_{j0j1}^{zz}$ for an $S$ = 7/2 72-site cluster with periodic boundary conditions (PBC) along the $a$ axis and OBC perpendicular to $a$. The GdZnPO interaction parameters $J_1$ $\sim$ $-$0.39 K, $J_2$ $\sim$ 0.57 K, and $D$ $\sim$ 0.30 K are applied. Zigzag-stripe correlations are evident in both $\xi_{j0j1}^{xy}$ and $\xi_{j0j1}^{zz}$ on the honeycomb lattice. To minimize boundary effects, the ground-state energy and order parameter were evaluated using the central hexagon consisting of six spins ($S_1$-$S_6$). For comparison, calculations were also performed on 15 hexagons marked by open circles [see (e) and (f)].}
	\label{figs4}
\end{figure}

Using the crystal structure of GdZnPO, we evaluated the dipolar interactions between the first-, second-, and third-nearest-neighbor Gd$^{3+}$ spins to be $J_1^\mathrm{d}$ = $-$0.04 K, $J_2^\mathrm{d}$ = 0.04 K, and $J_3^\mathrm{d}$ = 0.002 K, respectively. The magnitudes of $J_1^\mathrm{d}$ and $J_2^\mathrm{d}$ are about one order smaller than those of the experimentally determined exchange constants $J_1$ $\sim$ $-$0.39 K and $J_2$ $\sim$ 0.57 K~\cite{PhysRevLett.133.236704}, confirming that the dominant contributions to $J_1$ and $J_2$ arise from short-range superexchange interactions mediated via Gd-O-Gd pathways. For spin-spin couplings beyond second neighbors, direct Gd-O-Gd superexchange paths are absent, and the interatomic distances $|$Gd-Gd$|_{\geq3}$ $\geq$ 5.4 \AA~are considerably larger than those for $J_1$ ($|$Gd-Gd$|_1$ = 3.7 \AA) and $J_2$ ($|$Gd-Gd$|_2$ = 3.9 \AA). Owing to the highly localized nature of the Gd$^{3+}$ 4$f$ electrons, it is thus sufficient to consider only the first- and second-nearest-neighbor exchange interactions in the spin Hamiltonian [see Eq.~(\ref{eq1})]; longer-range interactions are negligible, being $\lesssim$ $J_3^\mathrm{d}$ $\sim$ 0.4\%$J_2$ in GdZnPO.

\begin{figure}
	\includegraphics[width=0.49\textwidth]{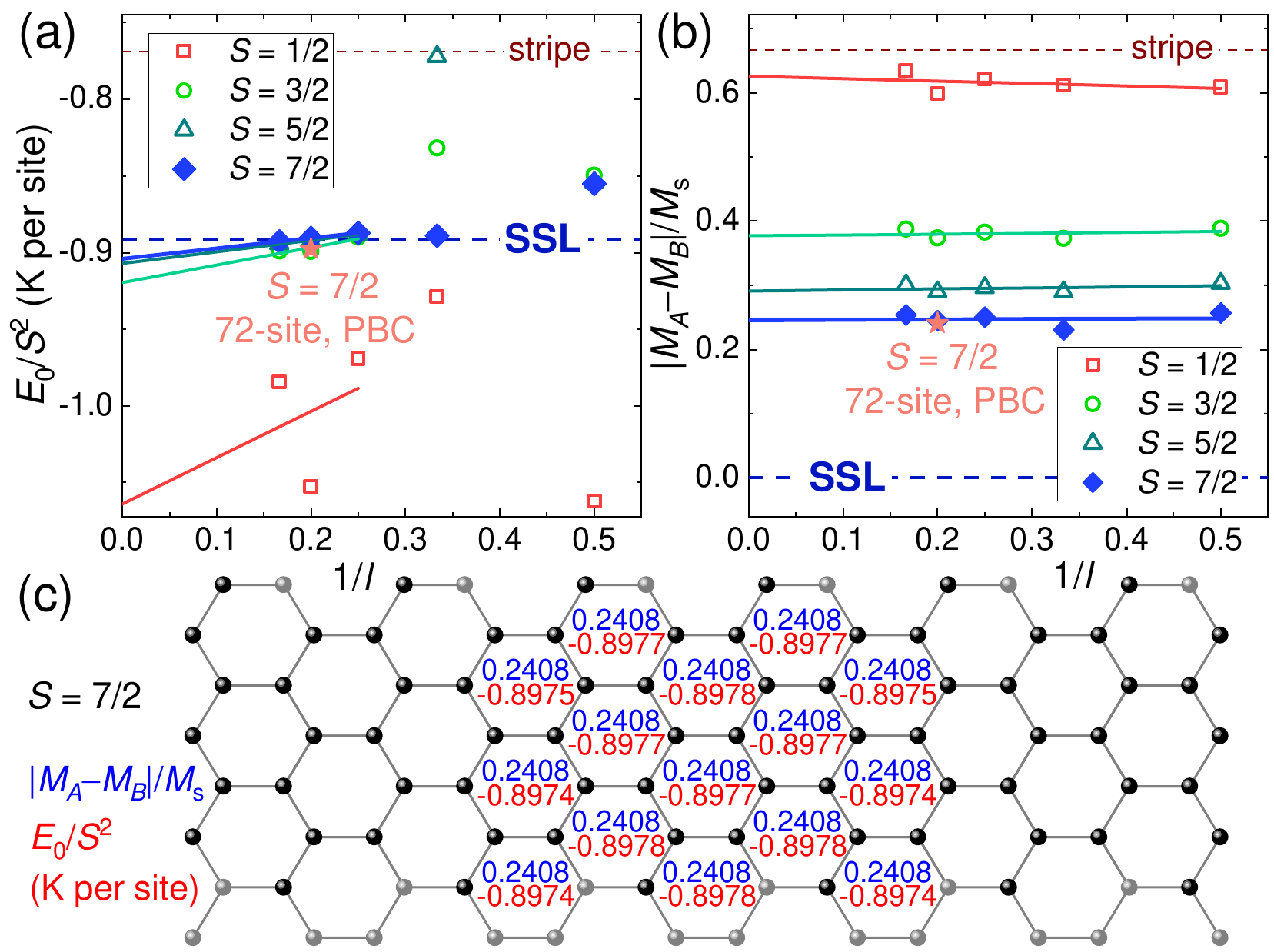}
	\caption{Finite-size scaling of ground-state properties. (a) Side-length ($l$) dependence of the ground-state energy per site ($E_0$) calculated for $S$ = 1/2, 3/2, 5/2, and 7/2 systems. The thick and thin dashed lines denote $E_0$/$S^2$ = $-$$\frac{J_1^2}{8J_2}$$-$$\frac{3}{2}J_2$ ($\sim-$0.89 K) and $\frac{1}{2}J_1$$-$$J_2$ ($\sim-$0.77 K), corresponding to the spiral spin liquid (SSL) and stripe state, respectively, in the classical ($S\to\infty$) limit. (b) Order parameters $|M_A$$-$$M_B|/M_\mathrm{s}$ calculated to compare with NMR observations. The thick and thin dashed lines represent $|M_A$$-$$M_B|/M_\mathrm{s}$ = 0 and 2/3 for the pure SSL and stripe (along $z$) state, respectively. Solid colored lines indicate linear extrapolations to the thermodynamic limit (1/$l$ $\to$ 0). Open boundary conditions [OBC; see Figs.~\ref{figs4}(a)-\ref{figs4}(d)] are used. Stars denote results for the $S$ = 7/2 72-site cluster with periodic boundary conditions (PBC) along $a$ and OBC perpendicular to $a$ [see (c)]. (c) Hexagon-dependent variation of the calculated $E_0$/$S^2$ and $|M_A$$-$$M_B|$/$M_\mathrm{s}$.}
	\label{figs5}
\end{figure}

The competing interactions $J_1$ and $J_2$ introduce spin frustration on the honeycomb lattice. At the classical level, the ground state of~Eq.~(\ref{eq1}) is expected to be an SSL~\cite{PhysRevB.106.035113}, with
\begin{multline}
\mathbf{S}_{j,A} =\\
S[\cos\psi\cos(\mathbf{Q}\cdot\mathbf{R}_j),\cos\psi\sin(\mathbf{Q}\cdot\mathbf{R}_j),\sin\psi]~\mathrm{and}
\label{eq5}
\end{multline}
\begin{multline}
\mathbf{S}_{j,B} =\\
S[\cos\psi\cos(\mathbf{Q}\cdot\mathbf{R}_j+\phi),\cos\psi\sin(\mathbf{Q}\cdot\mathbf{R}_j+\phi),\sin\psi],
\label{eq6}
\end{multline}
for spins residing on the triangular sublattices $A$ and $B$, respectively, where $\mathbf{Q}$ = $h\mathbf{b}_1$+$k\mathbf{b}_2$ is the wave vector, $\mathbf{b}_1$ and $\mathbf{b}_2$ are the reciprocal lattice vectors, and $\mathbf{R}_j$ denotes the position vector of the $j$th unit cell. Using this ansatz, the classical energy per site can be expressed as
\begin{multline}
E_\mathrm{class} = (D+\frac{3J_1}{2}+3J_2)S^2\sin^2\psi+h_\mathrm{Z}S\sin\psi\\
+\frac{J_1S^2\cos^2\psi}{2}[\cos\phi+\cos(\phi-2\pi h)+\cos(\phi-2\pi h-2\pi k)]\\
+J_2S^2\cos^2\psi[\cos(2\pi h)+\cos(2\pi k)+\cos(2\pi h+2\pi k)],
\label{eqs6}
\end{multline}
where $h_\mathrm{Z}$ = $-\mu_0Hg\mu_\mathrm{B}$. By minimizing~Eq.~(\ref{eqs6}), the angles $\phi$ and $\psi$ are determined as $\cos\phi$ = $\frac{F_1}{\sqrt{F_1^2+F_2^2}}$, $\sin\phi$ = $\frac{F_2}{\sqrt{F_1^2+F_2^2}}$, and $\sin\psi$ = $-\frac{h_\mathrm{Z}}{2S\eta}$. Here, $F_1$ = 1+$\cos(2\pi h_\mathrm{G})$+$\cos(2\pi h_\mathrm{G}$+$2\pi k_\mathrm{G})$, $F_2$ = $\sin(2\pi h_\mathrm{G})$+$\sin(2\pi h_\mathrm{G}$+$2\pi k_\mathrm{G})$, and $\eta$ = $D$+$\frac{3J_1}{2}$+$\frac{9J_2}{2}$+$\frac{J_1^2}{8J_2}$. Accordingly, the degenerate ``spiral contour" is determined by,
\begin{equation}
\cos(2\pi h_\mathrm{G})+\cos(2\pi k_\mathrm{G})+\cos(2\pi h_\mathrm{G}+2\pi k_\mathrm{G}) = \frac{1}{2}(\frac{J_1^2}{4J_2^2}-3),
\label{eqs7}
\end{equation}
demonstrating that it is entirely independent of the applied magnetic field. The minimum of~Eq.~(\ref{eqs6}), corresponding to the classical ground-state energy per site, is obtained as
\begin{equation}
E_0 = -\frac{h_\mathrm{Z}^2}{4\eta}-\frac{J_1^2S^2}{8J_2}-\frac{3}{2}J_2S^2\mathrm{ ~ ~ for~} H < H^*~\mathrm{or}
\label{eqs8}
\end{equation}
\begin{equation}
E_0 = (D+\frac{3J_1}{2}+3J_2)S^2+h_\mathrm{Z}S\mathrm{ ~ ~ for~} H \geq H^*,
\label{eqs9}
\end{equation}
where the crossover field is given by $\mu_0H^*$ = $S[2D$+$3J_1$+$9J_2$+$J_1^2/(4J_2)]/(g\mu_\mathrm{B})$ ($\sim$12 T). In the classical spiral ground state, the spins are uniformly polarized under a magnetic field applied along the $c$ axis, yielding $|M_A$$-$$M_B|$/$M_\mathrm{s}$ = 0.

\section{Quantum many-body simulations of ground-state properties}\label{AppendixC}

We performed DMRG calculations to investigate the ground-state properties of the GdZnPO spin Hamiltonian [Eq.~(\ref{eq1})] at zero magnetic field, employing $U$(1) symmetry, within \textsc{ITensor} software~\cite{Fishman2022The,Fishman2022Codebase}. As illustrated in Figs.~\ref{figs4}(a)-\ref{figs4}(d), several diamond-shaped clusters with open boundary conditions (OBC) were studied, with linear dimensions $l$ = 2-6 and a total of up to 98 spins. Up to 8000 density-matrix eigenstates were retained in the renormalization procedure, and approximately 20 sweeps were carried out until the ground-state energy ($E_0$/$S^2$) converged to within $\sim$ 6$\times$10$^{-5}$ K per site.

\begin{table}[h]
\caption{Nearest-neighbor ground-state correlations calculated using DMRG. To minimize boundary effects, $\xi_{j0j1}^{xy}$ and $\xi_{j0j1}^{zz}$ are listed for the six spins nearest the center of the $l$ = 5 cluster with open boundary conditions [see Figs.~\ref{figs4}(a)-\ref{figs4}(d)].}
\begin{tabular}{c|c|c|c|c}
\hline
\hline
  & \multicolumn{2}{c|}{$S$ = 1/2} & \multicolumn{2}{c}{$S$ = 7/2} \\
  \cline{2-3} \cline{4-5}
  & $\xi_{j0j1}^{xy}$ & $\xi_{j0j1}^{zz}$ & $\xi_{j0j1}^{xy}$ & $\xi_{j0j1}^{zz}$ \\
\hline
 $j0$ = 1, $j1$ = 2 & 0.24086 & 0.12295 & 0.56421 & 0.00127 \\
 $j0$ = 2, $j1$ = 3 & $-$0.15062 & $-$0.07793 & $-$0.87598 & 0.02173 \\
 $j0$ = 3, $j1$ = 4 & 0.28556 & 0.14534 & 0.80597 & $-$0.00339 \\
 $j0$ = 4, $j1$ = 5 & 0.24670 & 0.12572 & 0.56507 & 0.00097 \\
 $j0$ = 5, $j1$ = 6 & $-$0.15759 & $-$0.08107 & $-$0.87675 & 0.02190 \\
 $j0$ = 6, $j1$ = 1 & 0.28545 & 0.14510 & 0.80616 & $-$0.00363 \\
\hline
\hline
\end{tabular}
\label{tabs1}
\end{table}

Figures~\ref{figs4}(a) and~\ref{figs4}(b) presents the ground-state spin correlations $\xi_{j0j1}^{xy}$ = $\langle$GS$|S_{j0}^xS_{j1}^x$+$S_{j0}^yS_{j1}^y|$GS$\rangle$/$S^2$ and $\xi_{j0j1}^{zz}$ = $\langle$GS$|S_{j0}^zS_{j1}^z|$GS$\rangle$/$S^2$ obtained from DMRG calculations for $S$ = 1/2, revealing a clear zigzag-stripe pattern. Here, $|$GS$\rangle$ denotes the ground state. For $S$ = 1/2, the single-ion anisotropy terms in~Eq.~(\ref{eq1}) are constant and thus inactive, resulting in nearly isotropic spin correlations with $\xi_{j0j1}^{xy}$ $\sim$ 2$\xi_{j0j1}^{zz}$ [see Figs.~\ref{figs4}(a) and~\ref{figs4}(b), and Table~\ref{tabs1}]. In contrast, for $S$ $>$ 1/2, a pronounced easy-plane anisotropy emerges, characterized by $|\xi_{j0j1}^{xy}|$ $>$ 2$|\xi_{j0j1}^{zz}|$ [Figs.~\ref{figs4}(c) and~\ref{figs4}(d), and Table~\ref{tabs1}], which originates from the finite single-ion anisotropy $D$ $\sim$ 0.30 K ($>$ 0). As $S$ increases from 1/2 to 7/2, $|\xi_{j0j1}^{xy}|$ grows substantially but remains well below 1 even at $S$ = 7/2 (Table~\ref{tabs1}), indicating a significant deviation from the pure classical stripe order. Moreover, the correlations $\xi_{12}^{xy}$, $\xi_{23}^{xy}$, and $\xi_{34}^{xy}$ (or $\xi_{45}^{xy}$, $\xi_{56}^{xy}$, and $\xi_{61}^{xy}$) deviate from the values $\xi_0$, $-\xi_0$, and $\xi_0$, respectively, expected for a pure zigzag-stripe configuration, where 0 $<$ $\xi_0$ $\leq$ 1 is constant. These DMRG results support the possible coexistence of both spiral and zigzag-stripe correlations in the ground state of the spin-7/2 GdZnPO system at $\sim$0 T, consistent with the experimental findings.

To minimize boundary effects, we evaluated the ground-state energy per site ($E_0$) and the order parameter $|M_A$$-$$M_B|$/$M_\mathrm{s}$, which is directly related to the NMR shifts, using the six spins nearest to the cluster center (labeled 1-6 in Fig.~\ref{figs4}), as shown in Fig.~\ref{figs5},
\begin{multline}
E_0 = \langle\mathrm{GS}|\frac{J_1}{4}\sum_{\langle j0,j1=1\mathrm{-}6\rangle}\mathbf{S}_{j0}\cdot\mathbf{S}_{j1}+\frac{J_2}{2}\sum_{\langle\langle j0,j2=1\mathrm{-}6\rangle\rangle}\mathbf{S}_{j0}\cdot\mathbf{S}_{j2}\\
+\frac{D}{6}\sum_{j0=1\mathrm{-}6}(S_{j0}^z)^2|\mathrm{GS}\rangle,
\label{eqs10}
\end{multline}
\begin{multline}
|M_A-M_B|/M_\mathrm{s} =\\
\frac{\sqrt{\langle\mathrm{GS}|(\sum_{j0=1,3,5}S_{j0}^z-\sum_{j1=2,4,6}S_{j1}^z)^2|\mathrm{GS}\rangle}}{3S}.
\label{eqs11}
\end{multline}
To further mitigate finite-size effects, we performed linear fits to the large-cluster data and extrapolated both $E_0$/$S^2$ and $|M_A$$-$$M_B|$/$M_\mathrm{s}$ to the thermodynamic limit [$l$ $\to$ $\infty$, see Figs.~\ref{figs5}(a) and~\ref{figs5}(b)], as presented in the main text (see Fig.~\ref{fig4}).

Additionally, we performed DMRG calculations for a 72-site cluster with $S$ = 7/2, employing periodic boundary conditions along the $a$ axis and OBC perpendicular to it [Figs.~\ref{figs4}(e) and~\ref{figs4}(f)]. The calculations yield $E_0$/$S^2$ = $-$0.8977 K per site and $|M_A$$-$$M_B|$/$M_\mathrm{s}$ = 0.241, in good agreement with the OBC results [see Figs.~\ref{figs5}(a) and~\ref{figs5}(b)]. For the putative spiral or stripe orders on the honeycomb lattice, considering only the central six spins is theoretically sufficient to evaluate $E_0$ and $|M_A$$-$$M_B|$/$M_\mathrm{s}$ [Eqs.~(\ref{eqs10}) and (\ref{eqs11})]. Indeed, including the central 36 spins [Figs.~\ref{figs4}(e) and~\ref{figs4}(f)] gives $E_0$/$S^2$ = $-$0.8976 K per site and $|M_A$$-$$M_B|$/$M_\mathrm{s}$ = 0.241, in close agreement with the above six-spin results. Moreover, both quantities show no appreciable spatial variation across the central 15 hexagons in the cluster [see Fig.~\ref{figs5}(c)], validating the result.

Finally, the calculated order parameter $|M_A$$-$$M_B|$/$M_\mathrm{s}$ = 0.246(14) for $S$ = 7/2 remains larger than the experimental values of 0.11-0.16. This discrepancy likely arises from the limitations of conventional finite-size approaches in simulating SSL systems with incommensurate order~\cite{PhysRevLett.133.176502}. Accurately simulating strongly correlated quantum spin systems with incommensurate orders remains an outstanding challenge.

\end{document}